\def\Sone{S^{(1)}}
\def\Stwo{S^{(2)}}
\def\Soned{S^{(1)}_{diag}}
\def\Stwod{S^{(2)}_{diag}}
\def\deltam{{\delta_{max}}}
\def\Re{{\rm Re}}
\def\Im{{\rm Im}}
\def\imt{{\rm{Im}} T_{bb}}
\def\ret{{\rm{Re}} T_{bb}}
\def\meanxabsq{{{\overline{|X^2|}}}}
\def\meanxabsqp{{{\overline{|X'^{ 2}|}}}}
\def\meanxsq{{{\overline{X^2}}}}
\def\meanxstsq{{{\overline{X^{*2}}}}}
\def\to{\rightarrow}
\def\weight{{\cal W}}
\def\weightc{{\cal W}_{const}}
\def\weightw{{\cal W}_w}
\def\Dyson{{\cal D}}
\def\half{{\textstyle{1\over 2}}}
\def\begeq{\begin{equation}}
\def\begeqar{\begin{eqnarray}}
\def\endeqar{\end{eqnarray}}

\def\endeq{\end{equation}}
\def\cbar{{\overline c}}

\def\ubar{{\overline u}}

\def\kbar{{\overline K}}
\def\kbzero{{\overline K^0}}
\def\kbzerostar{{\overline K^{*0}}}
\def\konebar{{\overline K}_1}
\def\Bz{{B^0}}
\def\Bzbar{{\overline B}^0}
\def\Dzbar{{\overline D}^0}
\def\Dmbar{{D}^-}
\def\amp{{\cal{M}}}
\def\ortho{{\cal {O}}}

\def\ket#1{|#1>}

\def\sumxabs{{\textstyle\sum}|X^2|}
\def\sumxsq{{\textstyle\sum} X^2}
\def\sumxstsq{{\textstyle\sum} X^{*2}}
\def\xbabs{|X_b^2|}
\def\xbabsp{|X_{b'}^2|}
\def\sumxabs{{\textstyle\sum}|X^2|}

\def\Djm#1#2 {D^{#1}_{#2}}
\def\djm#1#2 {d^{#1}_{#2}}

\def\non{\nonumber\\}
\def\Dl#1#2{\delta_{#1#2}}

\def\triple#1#2#3#4#5#6{(#1#2)(#3#4)(#5#6)}
\def\double#1#2#3#4{(#1#2)(#3#4)}
\def\quadruple#1#2#3#4{(1#1)(2#2)(3#3)(4#4)}
\def\quadd#1#2#3#4#5#6#7#8{(#1#2)(#3#4)(#5#6)(#7#8)}
\input{psfig}
\documentstyle[12pt,amssymbols]{article}

\begin{document}
\begin{titlepage}
\begin{center}

\today \hfill    LBNL-40626\\
\hfill UCB-PTH-97/40\\
 
\vskip .2in

{\Large \bf Final-State Interactions in Nonleptonic Weak Decays of D and 
B Mesons}
\footnote{This 
work was supported in part by the Director, Office of 
Energy Research, Office of High Energy and Nuclear Physics, Division 
of 
High Energy Physics of the U.S. Department of Energy under Contract 
DE-AC03-76SF00098 and in part by the National Science Foundation under 
Grant PHY-95-14797.}

\vskip 0.15in
 
Robert N. Cahn
 
\vskip 0.1in
 
{\em Theoretical Physics Group\\
     Ernest Orlando Lawrence Berkeley National Laboratory\\
     University of California, Berkeley, California 94720}

\vskip 0.1in

Mahiko Suzuki

\vskip 0.1in

{\em Theoretical Physics Group\\
     Ernest Orlando Lawrence Berkeley National Laboratory\\
     University of California, Berkeley, California 94720}
 
and

{ \em Department of Physics\\
     University of California, Berkeley, California 94720}
 
\vskip 0.1in

\end{center}
 
\vskip .15in
 
\begin{abstract}
We study final-state interactions in nonleptonic weak decays in statistical 
models by averaging over ensembles of strong interaction S-matrices.  The models
range from one with completely random strong interactions, which gives 
extensive mixing between physical states, to models with feeble final-state
interactions, characterized by small phase shifts.  We compute
 expectation values 
for weak decay rates and the fluctuations around these means.  The
coherence between interfering amplitudes is gradually destroyed as the 
strength of the final-state interactions is increased.  Provided there 
are at least two weak phases, there is 
direct $CP$ violation as a result of the strong phases from final state 
interactions.  Data indicate that for $D$ meson decays, models with extensive 
mixing are appropriate, while for $B$ meson decays, models with small final
state interactions are required.
\end{abstract}
\end{titlepage}
\renewcommand{\thepage}{\roman{page}}
\setcounter{page}{2}
\mbox{ }
 
\vskip 1in
 
\begin{center}
{\bf Disclaimer}
\end{center}
\vskip .2in
\begin{scriptsize}
\begin{quotation}
This document was prepared as an account of work sponsored by the
United States Government. While this document is believed to contain
correct information, neither the United States Government nor any
agency thereof, nor The Regents of the University of California, nor
any of their employees, makes any warranty, express or implied, or
assumes any legal liability or responsibility for the accuracy,
completeness, or usefulness of any information, apparatus, product, or
process disclosed, or represents that its use would not infringe
privately owned rights.  Reference herein to any specific commercial
products process, or service by its trade name, trademark,
manufacturer, or otherwise, does not necessarily constitute or imply
its endorsement, recommendation, or favoring by the United States
Government or any agency thereof, or The Regents of the University of
California.  The views and opinions of authors expressed herein do not
necessarily state or reflect those of the United States Government or
any agency thereof, or The Regents of the University of California.
\end{quotation}
\end{scriptsize}
\vskip 2in
\begin{center}
\begin{small}
{\it Lawrence Berkeley Laboratory is an equal opportunity employer.}
\end{small}
\end{center}
\newpage
 
\renewcommand{\thepage}{\arabic{page}}
\setcounter{page}{1}
\setcounter{footnote}{0}
 
\section{Introduction}
\label{sec:intro}
\setcounter{equation}{0}
\setcounter{footnote}{0}
\setcounter{table}{0} 

The study of $CP$ violation is, to a large extent, the study of phases.  
Four-fermion interactions introduce $CP$  violation if their coefficients cannot
be made real through redefinition of the particle fields.  Kobayashi and 
Maskawa \cite{KM} recognized that, for three generations of quarks, mixing would
in general leave an irremovable phase, guaranteeing the presence of $CP$ 
violation, without the introduction of any new particles (aside from the $c$, 
$b$, and $t$ quarks, which were then yet to be discovered!).  It is not only 
through weak interactions, however, that phases are introduced into the 
amplitudes of weak decays.  When the final states 
include hadrons, and in 
nonleptonic decays in particular, there are strong final state interactions
that produce phases.  Strong phases arise in accordance with unitarity, which 
requires imaginary parts for amplitudes of processes that have on-shell physical 
intermediate states.  While the phase of a bare weak amplitudes changes sign
when the process is taken to its CP conjugate, the final-state interaction phase 
is unchanged, reflecting the CP invariance of strong interactions.

Final-state interaction phases can be either a blessing or a curse.  Some 
potential $CP$ violations are observable only if there are final state 
interactions present, with the measurable quantities being proportional,
typically, to the sine of the difference of two strong-interaction phases.  In 
other situations, the final-state interactions obscure the phases of the KM 
matrix that one wishes to measure.  Both circumstances demand understanding
of final-state interactions.

Unitarity and time-reversal invariance (of the strong interactions)
 determine final-state interaction phases. In the familiar case of
$K^0\to\pi\pi$ there are only two open channels, $I=0$ and $I=2$.  The weak
decays pick up phases $\exp(i\delta_I)$, where $\delta_I$ is the 
strong-interaction phase shift in the appropriate isospin channel.
Unfortunately, this simplicity is lost once there are many channels open to the 
strong interactions, as is the case for $D$ and $B$ decays.  Formally, it is
easy to describe the effect.  If there are $N$ open channels, we diagonalize the
S-matrix to find $N$ eigenchannels, each of which has an associated 
eigenphase-shift.  We then analyze the weak decays in this eigenbasis and 
attach the phase $\exp(i\delta_\alpha)$ to the eigenchannel $\ket\alpha$.

The difficulty here is two-fold.  We do not know a priori the eigenphase nor do
we know the composition of the eigenchannels.  These problems involve the 
full complexity of non-perturbative physics in a particularly awkward way.
In this paper, our approach is statistical, based on the assumption that when the number of
open channels, $N$, is large, it may be possible to study the general behavior
of the decay amplitudes, even when it is not possible to make a precise 
prediction for any particular one of them.  

How big are final-state interactions in the decays of heavy mesons?  
At one extreme, it might be argued that
they should be small.  An appeal can be made to asymptotic freedom, as has been
argued, for example, by Bjorken \cite{bj}.  An explicit implementation of this
philosophy was already given by Bander, Silverman, and Soni in 1979 
\cite{bander}, who calculated a one-loop diagram (a ``penguin'') that gave
an imaginary part to a decay amplitude, using perturbative QCD.  
At the other extreme, one can cite the highly inelastic nature of strong 
collisions at high energies and the success of models with absorption in low
partial waves\cite{jdj}.   

A crude but simple measure of the extent of final-state interactions is to 
look at decays to final states that involve interference of two
distinct isospin channels.  Consider, for example, $D^0\to K^-\pi^+, 
{\overline K}^0\pi^0$, whose final states are superpositions of $I=1/2$ and
$I=3/2$.  The ``factorization'' model argues that the charged current in weak
decays should produce a charged $\pi$ more frequently than neutral ones, by 
a factor 18 (even before QCD corrections, which further increase the 
dominance), requiring coherence 
between the $I=1/2$ and $I=3/2$ amplitudes.  In fact, the charged
 $\pi$ decay  has a branching ratio just a factor of two higher than the neutral
$\pi$ decay.  On the other hand, the decay $B^0\to D^-\rho^+$ is more than
an order of magnitude more probable than $B^0\to {\overline D}^0\rho^0$, 
consistent with factorization.  If
final-state interactions in the $I=1/2$ and $I=3/2$ channels were substantial
the relative phases between the $I=1/2$ and $I=3/2$ would have been randomized,
making a large discrepancy between the branching ratios improbable.  These
examples suggest that at the higher energies of $B$ decays, final-state
interactions are less important than at the lower energies of $D$ decays.  

Below we discuss a model that enables us to interpolate between the extremes of
complete final-state interaction mixing and the complete absence of such mixing.

Our point of departure is a statistical model of strong interactions based on
random S-matrices as introduced (initially for nuclear physics) by Dyson 
\cite{dyson}.  Dyson showed that each S-matrix can be expressed in terms of 
an $N$-dimensional orthogonal transformation (rotation) and a set of eigenphase shifts.  The weight
attached to each S-matrix is a product of the weight associated with the 
rotation group and a weight associated with the particular set of eigenphases.
This very elegant development is awkward to use, so we consider some 
simpler alternatives, as well.  The random S-matrices lead to nearly complete mixing of
all the communicating channels.  This may be appropriate to the $D$ decays, but
is not for the $B$ decays.  We develop an alternative in which the phase
shifts are restricted to a smaller range.
\section{Summary of Results}
\label{sec:results}
\setcounter{equation}{0}
\setcounter{footnote}{0}
\setcounter{table}{0}

The statistical approach of averaging over possible final-state interactions is
conceptually straightforward, but it does involve some tedious calculations, as
described in subsequent sections and as displayed more fully in some of the 
appendices.  The results of these calculations, however, are simple to grasp
and the essential points are displayed in this section.

We imagine a weak decay amplitude $X_b$ to a physical state $b$ in the absence of
strong interactions corrections.  The decay rate into $b$ is then, up to a fixed constant and 
phase space factors, 
\begeq
|\amp_b|^2=\xbabs.  
\endeq

In the presence of final-state interactions, a decay that ``initially'' produces
a state $a$ can rescatter into the state $b$.  The extent of this rescattering
can be modeled by introducing a distribution of phase shifts that are 
symmetrical about zero (so $<\sin\delta>=0$).  The strength of the phase 
shifts is characterized by $<\cos\delta>$ and $<\cos 2\delta>$.  In the extreme
of full final-state interactions, all values of the phase shift are equally
likely and  $<\cos\delta>=<\cos 2\delta>=0$.  In this instance, we find
\begeq
<|\amp_b|^2>={1\over N+2}(2\xbabs + N\meanxabsq),
\endeq
where
\begeq
\meanxabsq={1\over N}\sum_a|X_a^2|
\endeq
is the mean of the absolute squares of all the ``bare'' weak decay amplitudes.
For large $N$, most of the memory of the initial amplitude to $b$ is lost,
and instead it is only the average of the square of the amplitude that 
counts.  
More generally, if smaller values of the phase shifts are favored, we find
\begeqar
<|\amp_b|^2>&=&{N\over N+2}\left[{2\over N}|X_b|^2+\meanxabsq
   +\left(|X_b|^2-\meanxabsq\right)<\cos\delta>^2\right]\non
&=&{N\over N+2}\left[ (<\cos\delta>^2+{2\over N})|X_b|^2+(1-<\cos\delta>^2
)\meanxabsq\right].
\non\endeqar
We can see how this interpolates between the situation with no final-state
interactions, where only the amplitude to $b$ matters, and the fully mixed case.
We can study not only the expected means of the squares of the amplitudes, 
but also the fluctuations about those means. For large final state interactions,
the fluctuations are comparable to the mean squared amplitudes themselves.  
For small final-state interactions the fluctuations of course are small.

The coherence or incoherence of amplitudes for final states with different
isospin is an important feature of weak decays and reveals much about the decay
dynamics.  In the absence of final-state interactions, the actual decay 
amplitudes are simply the bare decay amplitudes and we can write

\begeq
<|\amp_b + \amp_{b'}|^2>=
 |X_b+X_{b'}|^2.\endeq
Here the primed and unprimed amplitudes correspond to two states that cannot
mix through strong interactions, because of differences in isospin or some
other good quantum number.
Introducing final-interactions mixes the amplitudes with other amplitudes, but
with the primed and unprimed groups remaining separate.  If there are $N$ 
unprimed states and $N'$ primed states, we find
\begeqar
&&<|\amp_b + \amp_{b'}|^2>\non
&&\quad = |<\cos\delta>X_b+<\cos\delta '>X_{b'}|^2\non
&&\qquad +{N(1-<\cos\delta>^2)\over N+2}
              \left({2\over N}\xbabs+\meanxabsq\right)\non
&&\qquad +{N'(1-<\cos\delta'>^2)\over N'+2}
              \left({2\over N'}\xbabsp+\meanxabsqp\right).\non
\endeqar

If the final-state interactions are not large, we can expand in the phase shifts
and obtain
\begeqar
<|\amp_b +& \amp_{b'}|^2>&\to
 |<\cos\delta>X_b +<\cos\delta'>X_{b'}|^2\non
&&\qquad   +<\delta^2>\meanxabsq +<\delta^{'2}>\meanxabsqp +{\cal O}(1/N)
\endeqar
We see that the rate receives contributions from a purely coherent piece and
from a purely incoherent piece, with the balance determined by the extent of
the final-state interactions.  There is complete coherence in their absence and
complete incoherence when the phase shifts are completely random.

Final-state interactions are essential to direct CP violation.  We find, to
leading order in $1/N$, that the expected size of such CP violation is 
given by 

\begeq
<\left(|{\amp}_{\overline b}^2|-|\amp_b^2|\right)^2>
={4\over N^2}\sum_{a,c}|X_a|^2|X_c|^2\sin^2 (\phi_a-\phi_c)
                 (1-<\cos 2\delta>^2).
\endeq
We see the expected structure.  CP violation requires at least two different
weak phases, $\phi_a, \phi_c,...$, 
together with some final state interactions.  If the phase shifts
are small, so are the direct violations of CP.

A comparison of some of the predictions with extant data on $D$ and $B$ decays
is given in Section \ref{sec:data}, to which the anxious reader can skip. 

\section{Unitarity and Final-State Interactions}
\label{sec:unitarity}
\setcounter{equation}{0}
\setcounter{footnote}{0}
\setcounter{table}{0} 

We are interested in decays of spinless particles so we restrict our consideration
to states of zero angular momentum.  It is conceptually simpler to think of
the final states as two-body so that they are discrete, say 
$N$ of them.  The
strong-interaction S-matrix in a fixed partial wave 
is thus $N\times N$ and in fact it can be taken to 
be symmetric, as a consequence of $T$ invariance.  
It satisfies the unitarity relation $SS^\dagger=I$.

The symmetric S-matrix can be diagonalized by an orthogonal transformation:

\begeq
S_{diag}=\ortho S \ortho^T,
\endeq
where $S_{diag}$ has diagonal elements $e^{2i\delta_\alpha}; \alpha=1,...N$.  
The real $N\times N$ matrices $\ortho$ satisfy

\begeqar
\ortho^{-1}&=&\ortho^T,\non
\sum_\alpha\ortho_{\alpha a}\ortho_{\alpha b}&=&\delta_{ab},\non
\sum_a\ortho_{\alpha a}\ortho_{\beta a}&=&\delta_{\alpha\beta},\non
\endeqar
where the Greek indices label states in the eigenbasis, while Latin indices
label physical states.

As shown in Appendix~\ref{appendixa}, the weak decay of a particle into the eigenstate 
$\alpha$
has an amplitude of the form 

\begeq
\amp_\alpha=e^{i\delta_\alpha}e^{i\phi_\alpha}W_\alpha,
\endeq
while the the CP conjugate decay has the amplitude

\begeq
\amp_{\overline\alpha}=e^{i\delta_\alpha}e^{-i\phi_\alpha}W_\alpha,
\endeq
where $W_\alpha$ is real,
that is, the weak phase, $\phi$, changes sign, while the strong phase, $\delta$,
does not.  The decay amplitude to a physical state, $b$, is related through
the orthogonal transformation, $\ortho$:

\begeqar
\amp_b&=&\sum_\alpha \ortho_{\alpha b}\amp_\alpha\non
&=&\sum_\alpha \ortho_{\alpha b}e^{i\delta_\alpha}e^{i\phi_\alpha}W_\alpha.
\endeqar
In particular

\begeqar
|\amp_b|^2&=&\sum_\alpha\ortho_{\alpha b}^2W_\alpha^2+\sum_{\alpha>\beta}
        2\ortho_{\alpha b}\ortho_{\beta b}
             \cos(\delta_\alpha+\phi_\alpha-\delta_\beta-\phi_\beta)
        W_\alpha W_\beta,\\
|\amp_{\overline b}|^2&=&\sum_\alpha\ortho_{\alpha b}^2W_\alpha^2+
       \sum_{\alpha>\beta}2\ortho_{\alpha b}\ortho_{\beta b}
             \cos(\delta_\alpha-\phi_\alpha-\delta_\beta+\phi_\beta)
                   W_\alpha W_\beta.
\endeqar
There is direct CP violation if
\begeq
|\amp_b|^2-|\amp_{\overline b}|^2=
      -4 \sum_{\alpha>\beta}\ortho_{\alpha b}\ortho_{\beta b}
             \sin(\delta_\alpha-\delta_\beta)\sin(\phi_\alpha-\phi_\beta)
                   W_\alpha W_\beta\label{eq2.8}
\endeq
does not vanish, which requires at least two strong channels,
with different phase shifts.  This could occur because two or more strong 
channels communicate, or because there are two non-communicating channels that
interfere (as
in the case of $K\to\pi\pi$).  In addition, there must be at least two 
distinct weak phases.  Once both these conditions are met, in general there
may be direct observable CP violation.  If we sum over all final states, $b$, the difference
in decay rates vanishes, as required by the CPT theorem.  

The strong-interaction matrix, $S$, has a number of invariant subspaces, that 
is, groups of states that don't get mixed
with states from outside the group.  Most obviously, isospin is a good quantum 
number so we can separate final states according to isospin.  Similarly, parity
and G-parity are good quantum numbers under strong interactions, 
and we can separate 
states by these, as well.  Consider, for example, $B^0$ decays.  The $\pi\pi$ 
final states appear in both the $I=0$ and $I=2$ subspaces.  The $\pi\pi$ final
states with $J=0$ have even parity and G-parity.   
The $\rho\pi$ final states are in 
different subspaces since they have odd parity and G-parity.
  The $\rho\rho$ final states have even G-parity and $I=0$ and $I=2$.
However, they can have either even parity (s-wave and d-wave decays) or 
odd parity (p-wave decays).  
If we sum the partial decay rates over all final states with a given set of 
good quantum numbers, it must equal the rate of the decay of the conjugate
particle into all final states with the CP-conjugate set of good quantum 
numbers.
Here we must allow for mixing between 
states like $D{\overline D}$ with $\pi\pi$, which although dynamically 
suppressed, is not forbidden through strong interactions.  

\section{An Ansatz}
\label{sec:ansatz}
\setcounter{equation}{0}
\setcounter{footnote}{0}
\setcounter{table}{0} 

When the theory of final state interactions was first introduced \cite{watson},
hadrons were considered elementary particles.  Nonleptonic decays could then
be viewed as a weak decay followed by hadronic final-state interactions.  It
was possible to imagine isolating the weak decay in the absence of strong
interactions.  This is no longer true for hadrons made of quarks, since the 
hadrons themselves are formed by the same forces responsible for the final-state
interactions.  

To describe nonleptonic decays within the context of QCD we first integrate 
out the short-distance effects, generating an effective weak interaction.  This
introduces no ambiguity.  However, it is still necessary to separate the 
part of the long-distance interaction responsible for the formation of hadrons,
from the remainder that causes the final-state interaction.  

Final-state interaction theory prescribes the form of the weak decay amplitudes
in the presence of strong interactions, but this, by itself, is not enough for
us to draw useful conclusions.  We adopt as our ansatz that the decay amplitude
$\amp_\alpha=e^{i\delta_\alpha}e^{i\phi_\alpha}W_\alpha$ is to be interpreted
as a weak amplitude $\amp_\alpha=e^{i\phi_\alpha}W_\alpha$ with a strong
final-state correction $e^{i\delta_\alpha}$.  The amplitude $W_\alpha$ includes
the physics of hadron formation, such as quark wave functions, as well as 
short-distance corrections to the weak interaction.  Since the phase factor
$e^{i\delta_\alpha}$ arises only from intermediate states that are on the 
mass shell, the off-mass-shell hadronic contributions are included implicitly
in $W_\alpha$.  With this interpretation, we can write the ``bare'' decay 
amplitude to the physical state $b$ in the absence of final-state interactions
as

\begeqar
X_b&=&\sum_\alpha \ortho_{\alpha b} e^{i\phi_\alpha}W_\alpha,\label{threeone}
\endeqar
absorbing the weak phase in the definition of $X_\alpha$.  Of course this is only
heuristic: In the absence of strong interactions we wouldn't even have the 
hadronic final state $b$.  Nonetheless, since we can write

\begeqar
e^{i\phi_\beta}W_\beta&=&\sum_b \ortho_{\beta b} X_b,
\endeqar
we can express the physical amplitude, $\amp$ in terms of the $X$ amplitudes
\begeqar
\amp_b&=&\sum_\alpha \ortho_{\alpha b}e^{i\delta_\alpha}
        \sum_a \ortho_{\alpha a}X_a\non
      &=&\sum_a\sqrt{S}_{ba}X_a,
\endeqar
where we have identified the ``square root of the S-matrix''
\begeq
\sqrt S_{ba}=\sum_\alpha \ortho_{\alpha b}e^{i\delta_\alpha} \ortho_{\alpha a}.
\endeq
{\em
We shall take as our 
ansatz that we imagine the $X_b$ to be fixed, while
we allow the strong interactions to vary}, that is, while we allow the 
matrices $\ortho$, and the phase shifts, $\delta_\alpha, \ldots$ to vary over
some range.  If the phase shifts are all set to zero, there is no final state
interaction and $\amp_b=X_b$.
The opposite extreme allows the phase shifts to vary as widely as possible.

The amplitude $X_b$ includes short-distance QCD corrections at 
energies much higher than the initial hadron mass, with the effective weak 
interactions improved by the renormalization group.  In general, final-state 
interactions not only introduce phrases, but modify magnitudes as well.  When
we vary the phases randomly with our Ansatz, leaving the $X_b$ fixed, we 
are ignoring the necessary correlation between phases and magnitudes.  This
is an inevitable shortcoming of our statistical approach.

\section{Dyson's Random S-Matrices}
\label{sec:dyson}
\setcounter{equation}{0}
\setcounter{footnote}{0}
\setcounter{table}{0} 

In 1962 Dyson proposed that in treating the distribution of resonance 
energy levels
in a heavy nucleus, it would be better to consider random S-matrices rather than
an ensemble of Hamiltonians.  Because the S-matrices are unitary and thus
bounded, they form
a compact set and it is possible to define an invariant measure on them.  The
invariant measure provides the proper weighting to make the S-matrices random.
Since the set of symmetric matrices do not form a group, finding the invariant
measure required some careful analysis.  Dyson was able to show that the 
proper weighting for a set of phase shifts (here we treat our ``square root of
the S-matrix'' as the equivalent of Dyson's S-matrix) is

\begeq
\Dyson=[2^{2N}\pi^N\Gamma(1+\half N)]^{-1}\prod_{i<j}|e^{i\delta_i}
          - e^{i\delta_j}|,
\endeq
which has unit normalization:
\begeq
\int \prod_k d\delta_k\Dyson=1.
\endeq

Altogether, the measure on the set of symmetric matrices is given by

\begeq
\mu(d\sqrt S)=\mu(dR) \prod_\gamma
d\delta_\gamma\Dyson,
\endeq
where $\mu(dR)$ is the measure on the $N$-dimensional orthogonal group.  We shall
never need the explicit form of $\mu(dR)$, but only the invariant nature of the
measure.

If we postulate that the $X_b$ are fixed, while varying the strong interactions
by integrating over the space of S-matrices (or rather their square roots),
we can compute ensemble averages, at least in principle.

Dyson's random S-matrices may be appropriate to decays in which final states
acquire phases that may be large, that is, in the resonance region.  The decays
of D mesons might fit this description.  When the final states are at higher
energies, beyond the resonance region, we expect that the phases shifts may be
small.  This might describe B decays.  
Below we consider Dyson's completely random matrices 
and a number of alternatives, and calculate
the observable quantities of interest for weak decays to see how final-state
interactions manifest themselves.

\section{Models for Final State Interactions}
\label{sec:models}
\setcounter{equation}{0}
\setcounter{footnote}{0}
\setcounter{table}{0} 

Dyson's construction of random S-matrices is a point of departure for 
phenomenological models.  We can generalize by replacing $\Dyson$ with some 
alternative weight, $\weight$.  The most obvious choice is the constant weight,
$\weightc=(2\pi)^{-N}$.  

As noted in the introduction, the persistence of a correlation in the phase
of different isospin amplitudes in $B$ decays (the success of factorization) 
indicates that a completely randomized strong-interaction description 
cannot be successful.  If the strong interaction effects become less 
important at higher energies, then a plausible description might be obtained
by restricting the range of the eigenphase-shifts.  We can model this 
behavior by selecting a function $w$ and defining

\begeq
\weightw=\prod_\gamma w(\delta_\gamma)
\endeq
with $w(-\delta)=w(\delta)$ and
\begeqar
\int_0^{2\pi}d\delta w(\delta)&=&1,\non
\int_0^{2\pi}d\delta w(\delta)\cos\delta&=&<\cos\delta>,\non
\int_0^{2\pi}d\delta w(\delta)\cos 2\delta&=&<\cos 2\delta>.\non
\endeqar
In the limit of small final-state interactions, $\delta$ is confined to a 
symmetrical region around $\delta=0$. At the other extreme, we can take $w=1/2\pi$
and recover the results for $\weight=\weightc$.  

We can get a sense of these approximations by calculating 
ratio of the expectation values,
averaged over the ensemble, of the elastic and total cross sections in the 
fixed zero angular momentum partial wave.
Writing $S=1+2iT$, we have

\begeq
{\sigma_{elastic}\over \sigma_{total}}={{\ret^2 +\imt^2}\over \imt}.
\endeq

Since the S-matrix can be written as

\begeq
S_{ba}=\sum_\alpha\ortho_{\alpha b}e^{2i\delta_\alpha}\ortho_{\alpha a},
\endeq
we have
\begeqar
\Re S_{bb}&=&
    \sum_\alpha\ortho_{\alpha b}\cos{2\delta_\alpha}\ortho_{\alpha b},\non
\Im S_{bb}&=&
    \sum_{\alpha}\ortho_{\alpha b}\sin{2\delta_\alpha}\ortho_{\alpha b},\non
(\Re S_{bb})^2&=&
    \sum_{\alpha,\beta}
         \ortho_{\alpha b}\cos{2\delta_\alpha}\ortho_{\alpha a}
            \ortho_{\beta b}\cos{2\delta_\beta}
                                   \ortho_{\beta b},\non
(\Im S_{bb})^2&=&
    \sum_{\alpha,\beta}
         \ortho_{\alpha b}\sin{2\delta_\alpha}\ortho_{\alpha a}
            \ortho_{\beta b}\sin{2\delta_\beta} \ortho_{\beta b}.\non
\endeqar

Thus, for example,
\begeqar
<(Re S_{bb})^2>&=&
\int \mu(dR)\prod_\gamma d\delta_\gamma \weight
               \sum_{\alpha,\beta}
         \ortho_{\alpha b}\cos{2\delta_\alpha}\ortho_{\alpha a}
           \ortho_{\beta b}\cos{2\delta_\beta} \ortho_{\beta b}.\non
\endeqar
\vfill\eject
Integrals over the $N$-dimensional orthogonal group
of products of $\ortho$s
appear throughout our calculations.  Results for up to eight $\ortho$s are 
derived in Appendix~\ref{appendixb}.  Using those results we find with $\weight=
\weightw$

\begeqar
<\Re S_{bb}>&=&<\cos 2\delta>,\non
<\Im S_{bb}>&=&<\sin 2\delta>=0,\non
<(\Re S_{bb})^2>&=&{3\over N+2}<\cos^22\delta>+{N-1\over N+2}<\cos 2\delta>^2
     , \non
<(\Im S_{bb})^2>&=&{3\over N+2}<\sin^22\delta> .     \non
\endeqar
Altogether, we find

\begeq
{<\sigma_{elastic}>\over< \sigma_{total}>}=<\sin^2\delta>+{3\over N+2}
          <\cos^2\delta>.\label{xsec}
\endeq
For completely random phase shifts, $<\sin^2\delta>=<\cos^2\delta>=1/2$, and
the elasticity is equal to one-half, up to corrections of order $1/N$.  The 
observed ratio of elastic to total cross sections 
in $\pi p$ and $K p$ scattering at $\sqrt s\approx 6
\ {\rm{GeV}}$ is about 15\%.  Of course, the scattering at these energies is 
due to many partial waves, but we should not be surprised to find a value of
$<\sin^2\delta>$ as small as 0.1 to 0.2 at the B-decay energy.

The average of the square of the decay matrix element over the ensemble of
unitary (square-root of) S-matrices is given by

\begeqar
<|\amp_b|^2>&= &\int \mu(dR) \int \prod d\delta_\gamma 
\weight
              \sum_a\left(\sqrt S_{ba}X_a\right)
                \sum_c\left(\sqrt S_{bc}X_c\right)^*\nonumber\\
            &= & \int \mu(dR) \int \prod d\delta_\gamma\weight
               \sum_{a,c}\sum_\alpha \ortho_{\alpha b}e^{i\delta_\alpha}
                                          \ortho_{\alpha a}
         \sum_\beta \ortho_{\beta b}e^{-i\delta_\beta}
                             \ortho_{\beta c} 
                              X_a X_c^*.\nonumber\\
\endeqar

The simplest alternative is $\weight=\weightc$, which has the virtue
$<\cos\delta>=0$.  We proceed to calculate for this choice
\begeqar
<|\amp_b|^2>
            &= & \int \mu(dR) \int \prod_\gamma {d\delta_\gamma\over 2\pi}
               \sum_{a,c}\sum_\alpha \ortho_{\alpha b}e^{i\delta_\alpha}
                                          \ortho_{\alpha a}
         \sum_\beta \ortho_{\beta b}e^{-i\delta_\beta}
                             \ortho_{\beta c} 
                              X_a X_c^*.\nonumber\\
\endeqar
Only terms with $\alpha=\beta$ contribute, so

\begeq
<|\amp_b|^2>= \int \mu(dR) \sum_{\alpha,a,c} \ortho_{\alpha b}\ortho_{\alpha a}
                       \ortho_{\alpha b}\ortho_{\alpha c}
                              X_a X_c^*.
\endeq 
Using the results from Appendix~\ref{appendixb}, we find, for $\weight=\weightc$,
\begeq
<|\amp_b|^2>={1\over N+2}(2\xbabs + N\meanxabsq),
\endeq
where
\begeq
\meanxabsq={1\over N}\sumxabs
\endeq
is the mean of the absolute squares of all the ``bare'' weak decay amplitudes.

We now turn our attention to 
calculating the fluctuations away from this 
result, i.e., we consider

\begeqar
<|\amp_b^2|^2>&=&\int \mu(dR)\int\prod d\delta_\gamma \weight\non
&&\times\sum_{a,c,d,e}\sum_{\alpha,\beta,\gamma,\epsilon}
          \ortho_{\alpha b} e^{i\delta_\alpha}\ortho_{\alpha a}
          \ortho_{\gamma b} e^{-i\delta_\gamma}\ortho_{\gamma c}
          \ortho_{\delta b} e^{i\delta_\delta}\ortho_{\delta d}
          \ortho_{\epsilon b} e^{-i\delta_\epsilon}\ortho_{\epsilon e}\nonumber\\
&&\qquad\times X_aX_c^*X_dX_e^*.
\endeqar
In evaluating this expression, we proceed by separating out categories with
varying number of Greek indices are identical, as shown in Table~\ref{t1} and
discussed further in Appendix~\ref{appendixc}.

If, as before, we set $\weight=\weightc$, we see that there 
are non-vanishing contributions only if $\alpha=\gamma$ and $\delta=\epsilon$ 
or if $\alpha=\epsilon$ and $\gamma=\delta$.  There are two kinds of 
contributions: ones where all four Greek indices are the same, and ones where 
there are two distinct Greek indices.  There are $N$ of the former and 
$2N(N-1)$ of the latter.  Thus the sum over eigenchannels we need in 
calculating the average over the orthogonal group is

\begeqar
&&\int \mu(dR)\left[ N(\ortho_{\alpha b} \ortho_{\alpha a} \ortho_{\alpha b} 
                \ortho_{\alpha c} \ortho_{\alpha b} \ortho_{\alpha d} 
                  \ortho_{\alpha b} \ortho_{\alpha e} )\right.\nonumber\\
&&\qquad        +2N(N-1)\left. (\ortho_{\alpha b} \ortho_{\alpha a} \ortho_{\alpha b} 
                \ortho_{\alpha c} \ortho_{\beta b} \ortho_{\beta d} 
                  \ortho_{\beta b} \ortho_{\beta e} )\right]
X_aX_c^*X_dX_e^*.\non\label{full}
\endeqar
\vfill\eject

\begin{table}[t]
\begin{center}
\begin{tabular}{|l|l|l|l|}\hline
Category&Condition&Terms&Phase\\ \hline
I&$\alpha=\gamma    =\delta=\epsilon$&$N$&$1$\\ \hline
II&$\alpha=\gamma\neq\delta=\epsilon$&$N(N-1)$&$1$\\ 
&$\alpha=\epsilon\neq\gamma=\delta$&$N(N-1)$&\\ \hline
III&$\alpha=\delta\neq\gamma=\epsilon$&$N(N-1)$&
         $e^{2i(\delta_\alpha-\delta_\gamma)}$\\ \hline
IV&$\alpha\neq\gamma\neq\delta\neq\epsilon$&$N(N-1)(N-2)(N-3)$&
         $e^{i(\delta_\alpha-\delta_\gamma+\delta_\delta-\delta_\epsilon)}$\\
     \hline
V&$\alpha\neq\gamma\neq\delta=\epsilon$&$N(N-1)(N-2)$&$e^{i(\delta_\alpha-
   \delta_\gamma)}$\\ 
&$\delta\neq\gamma\neq\alpha=\epsilon$&$N(N-1)(N-2)$&$e^{i(\delta_\delta-
   \delta_\gamma)}$\\ 
&$\delta=\gamma\neq\alpha\neq\epsilon$&$N(N-1)(N-2)$&$e^{i(\delta_\alpha-
   \delta_\epsilon)}$\\ 
&$\alpha=\gamma\neq\delta\neq\epsilon$&$N(N-1)(N-2)$&$e^{i(\delta_\delta-
   \delta_\epsilon)}$\\ \hline
VI&$\alpha\neq\delta\neq\gamma=\epsilon$&$N(N-1)(N-2)$&$e^{i(\delta_\alpha
+   \delta_\gamma-2\delta_\epsilon)}$\\
&$\alpha =\delta\neq\gamma\neq\epsilon$&$N(N-1)(N-2)$&$e^{i(2\delta_\alpha-
   \delta_\gamma-\delta_\epsilon)}$\\ \hline
VII&$\alpha=\gamma=\delta\neq\epsilon$&$N(N-1)$&$e^{i(\delta_\delta -
   \delta_\epsilon)}$\\ 
 &$\alpha=\delta=\epsilon\neq\gamma$&$N(N-1)$&$e^{i(\delta_\alpha-
   \delta_\gamma)}$\\ 
&$\gamma=\delta=\epsilon\neq\alpha$&$N(N-1)$&$e^{i(\delta_\alpha -
   \delta_\gamma)}$\\ 
 &$\alpha=\gamma=\epsilon\neq\delta$&$N(N-1)$&$e^{i(\delta_\delta -
   \delta_\epsilon)}$\\ \hline
\end{tabular}\end{center}
\caption[a]{Various categories of integrals over the $N$-dimensional orthogonal
group of products of eight $\ortho$s.  Since each Greek index can take on any
of $N$ values, the total number of terms is $N^4$.  The integral over the phase
depends on the choice of $\weight$.  The result of the integral over the
rotations is a tensor, which is contracted with $X_aX_c^*X_dX_e^*$.
\label{t1}}
\end{table}
Using the results of Appendix~\ref{appendixb}, and writing
\begeqar
\meanxsq&=&{1\over N}\sumxsq,\non
\meanxstsq&=&{1\over N}\sumxstsq.
\endeqar
for $\weight=\weightc$,
\begeqar
<|\amp_b^2|^2>&=&{1\over (N+1)(N+2)(N+4)(N+6)}\nonumber\\
&&\quad \times \left[2(N+3)(N+5)N^2(\meanxabsq)^2+8N^2(N+5)\meanxabsq\xbabs
  \right.\nonumber\\
&&\quad\qquad -4N(N+3)(\meanxsq X_b^{*2}+\meanxstsq X_b^2)+8(N^2+7N+9)\xbabs^2
      \nonumber\\ &&\qquad\qquad\left.+N^2(N+9)\meanxsq
      \meanxstsq\right].
\nonumber\\
\endeqar
To leading order in $1/N$, the fluctuations are given by
\begeqar
<|\amp_b^2|^2>-<|\amp_b|^2>^2 
&\approx&
             \left(\meanxabsq +{2\over N}\xbabs\right)^2\non
&\approx&   <|\amp_b|^2>^2.\label{eq518} \non
\endeqar
It is not much harder to compute for the more general case $\weight=\weightw$.
We find
\begeqar
<|\amp_b|^2>&=&{N\over N+2}\left[{2\over N}|X_b|^2+\meanxabsq
   +\left(|X_b|^2-\meanxabsq\right)<\cos\delta>^2\right]\non
&=&{N\over N+2}\left[ (<\cos\delta>^2+{2\over N})|X_b|^2+(1-<\cos\delta>^2
)\meanxabsq\right].\label{eq5.19}
\non\endeqar
We recover the results for $\weight=\weightc$  by setting $<\cos\delta>=0$.  
In the limit of small phase shifts, we see that the effect of rescattering
vanishes.
 
A tedious calculation employing extensively the results of 
Appendix~\ref{appendixb} yields
the fluctuations in this result.  To leading order in $1/N$, we find,

\begeqar
&&<|\amp_b^2|^2>-<|\amp_b|^2>^2\non
&& \approx
\meanxabsq^2\left[(1-<\cos\delta>^2)^2+(<\cos\delta>^2
                        -<\cos2\delta>)^2\right]\non
&&\qquad +|X_b|^2\meanxabsq 
                   \left[ 2<\cos\delta>^2 -4<\cos\delta>^4
                             +2<\cos\delta>^2<\cos 2\delta>\right]\non
&&\qquad +\xbabs^2 {1\over N}\left[4<\cos\delta>^2-8<\cos\delta>^4+4<\cos\delta>^2
                              <\cos 2\delta>\right].\non
\endeqar
The fluctuations vanish, of course, in the limit of zero phase shift.  In
addition, the term varying as $|X_b^4|$ is suppressed by $1/N$.
If the phase shifts are small,
\begeq
<|\amp_b|^2>\approx |X_b|^2 +<\delta^2>\meanxabsq
\endeq
and the fluctuations are

\begeqar
<|\amp_b^2|^2>-<|\amp_b|^2>^2 &\approx&
2<\delta^2>^2\meanxabsq^2+(<\delta^4>-<\delta^2>^2)(\xbabs\meanxabsq+{2\over N}
     \xbabs^2).\non
\endeqar

We turn now to the calculation with the full Dyson weight, $\weight=\Dyson$

\begeq
<|\amp_b|^2>
            =\int \mu(dR) \int \prod d\delta_k\Dyson
               \sum_{a,c}\sum_\alpha \ortho_{\alpha b}e^{i\delta_\alpha}
                                          \ortho_{\alpha a}
         \sum_\gamma \ortho_{\gamma b}e^{-i\delta_\gamma}
                             \ortho_{\gamma c} 
                              X_a X_c^*.
\endeq
We find after integrating over the $\ortho$s:

\begeq
<|\amp_b|^2>
            =\int \prod d\delta_k\Dyson {N\over N+2}
        \left[{2\over N}|X_b|^2 + \meanxabsq + \cos(\delta_1-\delta_2)(\xbabs 
                   - \meanxabsq)\right].\label{meansq}
\endeq
In \cite{dysoniii}, Dyson shows how to calculate the n-point correlation 
function:
\begeq
R_n(\delta_1,...\delta_n)={N!\over (N-n)!}\int\prod_{k>n} d\delta_k\Dyson.
\endeq
It is 
immediate that
\begeqar
R_0&=&1,\nonumber\\
R_1(\delta)&=&{N\over 2\pi}.
\endeqar
In Appendix~\ref{appendixd} we show that 
\begeqar
\int\prod_{k} d\delta_k\Dyson \cos(\delta_1-\delta_2)&=&-{N^2+1
 \over (N-1)(N^2-1)}.
\endeqar
Inserting this value, we find
\begeq
<|\amp_b|^2>
            = {N\over N+2}
        \left[{2\over N}|X_b|^2 + \meanxabsq - {N^2+1\over (N^2-1)(N-1)}(\xbabs 
                   - \meanxabsq)\right].
\endeq

It is apparent that the inclusion of the Dyson factor induces only a small 
change, one which vanishes if the bare amplitude has the same magnitude as
the average amplitude, and which is suppressed by one power of $1/N$.

\section{Coherence of Interfering Amplitudes}
\label{sec:coherence}
\setcounter{equation}{0}
\setcounter{footnote}{0}
\setcounter{table}{0} 

Coherence of interfering amplitudes is a measure of final state interactions.
We can address this question quite simply using the weight $\weightw$.
Suppose we have two interfering amplitudes from 
distinct channels (e.g., $I=1/2$ and $I=3/2$) so that the full amplitude
is 

\begeq
\amp_b +\amp_{b'}=\sum_{a,\alpha}\ortho_{\alpha b}e^{i\delta_\alpha}
                    \ortho_{\alpha a}X_a +
\sum_{a',\alpha '}\ortho'_{\alpha ' b'}e^{i\delta_\alpha '}
                    \ortho'_{\alpha 'a'}X_{a'}. 
\endeq
When we average over the ensemble of S-matrices we average over the two sectors
containing $b$ and $b'$, which are completely independent.  We thus find

\begeqar
&&<|\amp_b + \amp_{b'}|^2>\non
&&\quad = {N\over N+2}\left[(<\cos\delta>^2+{2\over N})\xbabs 
                             +(1-<\cos\delta>^2)\meanxabsq\right]\non
&&\qquad +{N'\over N'+2}\left[(<\cos\delta'>^2+{2\over N'})\xbabsp 
                             +(1-<\cos\delta'>^2)\meanxabsqp\right]\non
&&\qquad \quad+2 {\rm Re}X_bX_{b'}^*<\cos\delta><\cos\delta '>.\non
\endeqar

If the phase shifts are small, coherence is maintained as is seen from rewriting
this as 

\begeqar
&&<|\amp_b + \amp_{b'}|^2>\non
&&\quad = |<\cos\delta>X_b+<\cos\delta '>X_{b'}|^2\non
&&\qquad +{N(1-<\cos\delta>^2)\over N+2}
              \left({2\over N}\xbabs+\meanxabsq\right)\non
&&\qquad +{N'(1-<\cos\delta'>^2)\over N'+2}
              \left({2\over N'}\xbabsp+\meanxabsqp\right).\non\label{eq6.3}
\endeqar

When the phase shifts vary over a broad range so that 
$<\cos\delta>\approx 0$, the right hand side  of Eq. (\ref{eq6.3})
gives simply the incoherent sum:
\begeq
<|\amp_b + \amp_{b'}|^2>\to <|\amp_b|^2>+<|\amp_{b'}|^2>;
\qquad<\cos\delta>\to 0,
\endeq
and there is no observed interference.  In contrast, when the final state 
interactions are small and $<\delta^2>\to 0,$
\begeqar
<|\amp_b +& \amp_{b'}|^2>&\to
 |<\cos\delta>X_b +<\cos\delta'>X_{b'}|^2\non
&&\qquad   +<\delta^2>\meanxabsq +<\delta^{'2}>\meanxabsqp \non
&&\qquad\qquad+{\cal O}(1/N)
;\qquad
         <\delta^2>\to 0.
\endeqar

Numerically, the incoherence introduced by 
final state interactions in this case turns out to be 
surprisingly small.  It is the feature that makes the factorization hypothesis
work in the B decays, with their small final state interactions, whereas 
factorization fails in the D decays, where the final state interactions are
substantial.  A more quantitative treatment of these issues is given in Section
\ref{sec:data}.

\section{Direct CP Violation}
\label{sec:directcp}
\setcounter{equation}{0}
\setcounter{footnote}{0}
\setcounter{table}{0} 

As discussed in Section 2, direct CP violation is observable when the $X_a$ are
complex, provided there are final state interactions and more than one weak
phase.  The rates for a decay and its CP conjugate are given by

\begeqar
<|\amp_b|^2>
            &=& \int \mu(dR) \int \prod d\delta_k \weight
               \sum_{a,c}\sum_\alpha \ortho_{\alpha b}e^{i\delta_\alpha}
                                          \ortho_{\alpha a}
         \sum_\gamma \ortho_{\gamma b}e^{-i\delta_\gamma}
                             \ortho_{\gamma c} 
                              X_a X_c^*,\nonumber\\
<|{\amp}_{\overline b}|^2>
            &= & \int \mu(dR) \int \prod d\delta_k\weight
               \sum_{a,c}\sum_\alpha \ortho_{\alpha b}e^{i\delta_\alpha}
                                          \ortho_{\alpha a}
         \sum_\gamma \ortho_{\gamma b}e^{-i\delta_\gamma}
                             \ortho_{\gamma c} 
                              X_a^* X_c. \nonumber\\
\endeqar

These have equal expectations provided there is no a priori distinction between
the channels, i.e. so long as the weight $\weight(\delta_1, \delta_2, \ldots
\delta_N)$ is symmetric under the interchange of any two $\delta$s.  This is
true of all the models we consider and is implicit in our statistical approach
in general.  Of course, any particular choice of S-matrix will introduce 
direct CP violation, but the effect averaged over any ensemble
symmetric in the $\delta$s will vanish.  In particular, we can write

\begeqar
<|{\amp}_{\overline b}|^2>
            &=& \int \mu(dR) \int \prod d\delta_k\weight
               \sum_{a,c}\sum_\alpha \ortho_{\alpha b}e^{-i\delta_\alpha}
                                          \ortho_{\alpha a}
         \sum_\gamma \ortho_{\gamma b}e^{i\delta_\gamma}
                             \ortho_{\gamma c} 
                              X_a X_c^*.\nonumber\\
\endeqar

Consider next 

\begeqar
&&<\left(|{\amp}_{\overline b}^2|-|\amp_b^2|\right)^2>\non
&&\qquad      =\int \mu(dR) \int \prod d\delta_k\weight
\sum_{a,b,c,d}\sum_{\alpha,\beta,\gamma,\epsilon}
          \ortho_{\alpha b}\ortho_{\alpha a}
          \ortho_{\gamma b}\ortho_{\gamma c}
          \ortho_{\delta b}\ortho_{\delta d}
          \ortho_{\epsilon b}\ortho_{\epsilon e}\non
&&\qquad\times 2e^{i(\delta_\alpha-\delta_\gamma)}
   (e^{i(\delta_\delta-\delta_\epsilon)}-e^{-i(\delta_\delta-\delta_\epsilon)})
         X_a X_c^* X_d X_e^*.\label{directcp}
\endeqar
The effect of the second term in the bracket is to change the phase in a particular way.  
Referring to Table~\ref{t1} and Appendix~\ref{appendixc}, we see that the phase change interchanges the 
second instance in Category II with Category III, and the second and third 
instances of Category V to Category VI.  To leading order in $N$,
then, the result is, for $\weight=\weightw$,

\begeqar
<\left(|{\amp}_{\overline b}^2|-|\amp_b^2|\right)^2>
        &=&2   ( \meanxabsq-\meanxsq\meanxstsq)
                 (1-<\cos 2\delta>^2)\non
&=&{4\over N^2}\sum_{a,c}|X_a|^2|X_c|^2\sin^2 (\phi_a-\phi_c)
                 (1-<\cos 2\delta>^2).\non
\endeqar
This is to be compared to Eq.(\ref{eq5.19}) taken in the same limit, 

\begeq
<|\amp_b|^2>= \meanxabsq +<\cos\delta>^2(\xbabs-\meanxsq).
\endeq

With large final state interactions so $<\cos\delta>=0$, the relative 
magnitude of CP violation is

\begeqar
{<\left(|{\amp}_{\overline b}^2|-|\amp_b^2|\right)^2>\over
<\left(|{\amp}_{\overline b}^2|+|\amp_b^2|\right)^2>}
&=&{\sum_{a,c}|X_a|^2|X_c|^2\sin^2 (\phi_a-\phi_c)\over
              N^2  \meanxsq^2}.\non
\endeqar
When the phase shifts are small, $<\delta^2>\ll 1$, the relative magnitude is

\begeqar
{<\left(|{\amp}_{\overline b}^2|-|\amp_b^2|\right)^2>\over
<\left(|{\amp}_{\overline b}^2|+|\amp_b^2|\right)^2>}
&=&{4<\delta^2>\sum_{a,c}|X_a|^2|X_c|^2\sin^2 (\phi_a-\phi_c)\over
                N^2\xbabs^2},\non
\endeqar
which, naturally, is suppressed by the small strong phases, as anticipated in
Eq. (\ref{eq2.8}).
\section{Relation to Quark-level Analysis}
\label{sec:quark-level}

The short-distance portion of final state interactions has been computed in
perturbative QCD.  The result has been applied to CP violation phenomena
in B decays using some models for the soft strong interactions 
\cite{bander,gerard,palmer}.   Our statistical approach is complementary to
this QCD based model.  We formulate final state interactions solely in terms
of states of hadrons, filling in for our ignorance of detailed interactions with
the assumption of statistical randomness.  Rather than trying to predict the 
precise value of a process we attempt to estimate an ensemble average, the
ensemble being a range of strong interactions, as characterized by their 
S-matrices.

In the quark-level approach, the effective weak interaction Hamiltonian
is written

\begeq
{\cal H}=\sum_i c_i(\mu) {\cal O}_i,\label{op}
\endeq
where $c_i(\mu)$ incorporates the short-distance QCD loop corrections, and 
the ${\cal O}_i$ are four-quark operators that carry weak CP-violation phases.
The coefficients $c_i(\mu)$ are computed by the renormalization group method 
with $\mu^2$ kept in the spacelike region.  Therefore the $c_i(\mu)$ are real
and thus provide no final-state-interaction phases.  If the matrix elements of
the operators $\ortho_i$ are computed allowing one-loop corrections, imaginary
parts can arise.  The one-loop corrections of the matrix elements combine with
corrections to the coefficient functions so that the result may be written
\begeq
{\cal H}=\sum_i c_i^{eff}(\mu) {\cal O}_i^{tree},
\endeq
where now the $c_i^{eff}$ do carry strong interaction phases.  
Though the one-loop corrections to the matrix elements produce strong 
interaction phases for all interactions $\ortho_i$, Ref. \cite{palmer} argues
that the dominant phases arise from matrix elements of penguin-type operators,
the source identified in Ref. \cite{bander}.

This procedure generates final-state interactions from short-distance phenomena.
The formation of the hadrons itself introduces no phases, since it is carried
out using models \cite{bsw} based on factorization.  The accuracy of this
perturbative calculation of strong phases is yet to be tested, but it has been
advocated on the basis of asymptotic freedom \cite{bj}.  

In the quark-based approach, the strong phase of $<b|{\cal H}_{eff}|B>$ arises
from the effective coefficient functions of the various four-quark operators.  
Because
different final states receive different contributions from tree 
operators, their 
phases, which are generated by the coefficients of these operators, will vary.

A rough comparison can be made with our approach.  We might imagine the
effective Hamilton of Eq.(\ref{op}) to give our amplitude without final-state
interactions, i.e. Eq.(\ref{threeone}).  In place of the perturbative calculation
of hadronic rescattering, our procedure introduces the eigenphases of the 
S-matrix, which are, of course, unknown.  This results in a loss of explicit
predictive power, but provides a convenient formalism for discussing 
final-state interactions.  

\section{Comparison with Experiment}
\label{sec:data}
We have argued that a purely random S-matrix, with the phase shift taking on 
all possible values with equal likelihood would work best at energies 
where the strong
interactions produced resonances, while above those energies the phase shifts
would be confined to relatively small values.  Thus in the decay of D mesons, 
with mass a bit below 2 GeV, the random model might apply, while for B meson
decays, the final state interactions should be less substantial.  We consider
D and B decays in light of these expectations.

\subsection{D decays}

The development of the preceding Sections images all decays to be into two-body
final states.  Even broadening this to include quasi-two-body states cannot
account for all decays.  How then are we to determine $N$, the number of 
discrete final states?  We work empirically, inferring an effective channel
number that acts like $N$.   Fortunately, our predictions do not depend crucially
on the value of $N$, only on its being large.

We estimate $N$ by looking at the observed branching ratios.  The fully random
model predicts that the squares of the matrix elements should cluster around the 
mean
\begeq
<|\amp_b|^2>\approx \meanxabsq
\endeq
with the standard deviation equal to the mean itself.  Continuing to neglect
phase space corrections, we can relate the mean branching ratio to the inclusive
branching ratio to the category considered.  Thus the effective number of 
decay channels for non-leptonic D decays is roughly
\begeq
N={{\rm BR(inclusive)}_{non-leptonic}\over {\overline{BR}}_{non-leptonic}}
\endeq
Here we ignore the Cabibbo-suppressed decays.  We must, however, separate the 
$0^+$ and $0^-$ states since they are not mixed by the strong interactions.
The relevant branching ratios taken from the Review of Particle Properties 
\cite{rpp} are
listed in Tables \ref{plus} and \ref{zero}.

\begin{table}[htb]\begin{center}
\begin{tabular}{|l|r||l|r|}\hline
$D^+\ $ &BR in \%
&$D^+\ $&BR in \%\\ 
$J^P=0^+$ final state&
&$J^P=0^-$ final state&\\ \hline
$\kbzero\pi^+ $&$2.74\pm0.29$&$\kbzero \rho^+ $& $6.6\pm2.5$\\ \hline
$\kbzero a_1^+ $& $8.1\pm1.7$&$\kbzero a_2^+$&$<0.3$\\ \hline
$\kbzerostar \rho^+ $ &$2.1\pm1.4$&$\kbzerostar\pi^+$&$1.92\pm0.19$\\ \hline
$\konebar(1270)^0\pi^+$&$<0.7$&$\kbar(1410)^{*0}\pi^+$&$<0.7$\\ \hline
$\konebar(1400)^0\pi^+$&$5.0\pm1.3$&$\kbar_0(1430)^{*0}\pi^+$& $3.7\pm0.4$\\ \hline
&&$\kbar(1680)^{*0}\pi^+      $          &        $1.45\pm0.31$\\ \hline
\end{tabular}
\caption[a]{Branching ratios of $D^+$ into states with $J^P=0^+$ and $J^P=
0^-$ taken from
Ref. \cite{rpp}.  Final state that can be either $0^+$ or $0^-$ are also
included under $0^+$.   The mean branching ratio for $0^+$ is $3.6\pm0.9\%$. 
while for $0^-$ it is $2.3\pm0.6\%$.}
\label{plus}\end{center}
\end{table}

\begin{table}[htb]\begin{center}
\begin{tabular}{|l|r||l|r|}\hline
$D^0\ $&BR in \%
 &$D^0\ $&BR in \%\\ 
$J^P=0^+$ final state&
 &$J^P=0^-$ final state&\\ \hline
$K^-\pi^+      $          &        $3.83\pm0.12$
&$K^-\rho^+      $          &        $10.8\pm1.0$\\ \hline
$\kbzero \pi^0     $          &        $2.11\pm0.21$
&$\kbzero \rho^0     $          &        $1.20\pm0.17$\\ \hline
$\kbzero \eta      $          &        $0.7\pm0.1$
&$\kbzero \omega     $          &        $2.1\pm0.4$\\ \hline
$\kbzero \eta'      $          &        $1.70\pm0.26$
&$\kbzero f_0(980)      $          &        $0.57\pm0.16$\\ \hline
$K^-a_1^+      $          &        $7.3\pm1.1$
&$\kbzero \phi      $          &        $0.85\pm 0.10$\\ \hline
$\kbzero a_1^0      $          &        $<1.9$
&$\kbzero f_2(1270)      $          &        $0.41\pm 0.15$\\ \hline
$\kbzerostar \rho^0      $          &        $1.47\pm0.33$
&$K^{-} a_2(1320)^+      $          &        $<0.2$\\ \hline
$K^{*-}\rho^+      $          &        $6.0\pm2.4$
&$K^{*-}\pi^+      $          &        $5.0\pm0.4$\\ \hline
$\konebar(1270)^-\pi^+      $          &        $1.06\pm0.29$
&$\kbzerostar\pi^0      $          &        $3.1\pm0.4$\\ \hline
$\konebar(1400)^-\pi^+      $          &        $<1.2$
&$\kbar(1410)^{*-}\pi^+      $          &        $<1.2$\\ \hline
$\konebar(1400)^0\pi^0      $          &        $<3.7$
&$\kbar_0(1430)^{*-}\pi^+      $          &        $1.24\pm0.26$\\ \hline
$K^{*0}\omega      $          &        $1.1\pm0.4$
&$\kbar_2(1430)^{*-}\pi^+      $          &        $<0.8$\\ \hline
&&$\kbar(1410)^{*0}\pi^0      $          &        $<0.4$\\ \hline
&&$K^{*0}\eta        $          &        $1.9\pm0.5$\\ \hline
&&$K^{*0}\eta'       $          &        $<0.11  	 $\\ \hline
\end{tabular}
\caption[a]{Branching ratios of $D^0$ into states with $J^P=0^+$ and 
$J^P=0^-$ taken from
Ref. \cite{rpp}.  Final states that can be either $0^+$ or $0^-$ are also
included under $J^P=0^+$.   The mean branching ratio 
for $J^P=0^+$ is $2.1\pm0.4\%$, while for $J^P=0^-$
it is $1.7\pm 0.2$.}
\label{zero}\end{center}
\end{table}

\begin{figure}[htb]
\centerline{\psfig{figure=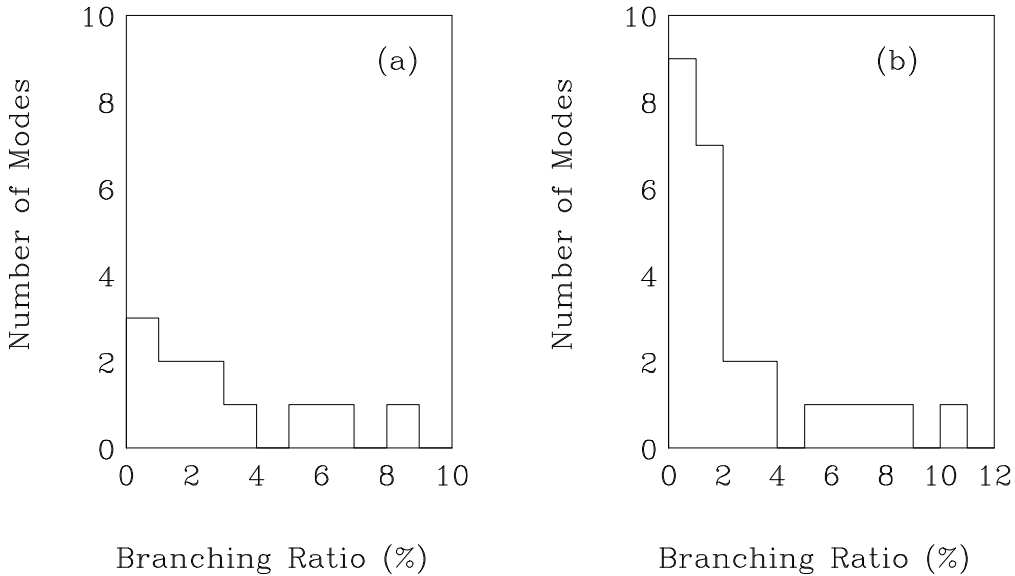,angle=90,width=5.25in}}
\caption[a]{(a) Decay histograms for $D^+$ non-leptonic decays, combining both
$J^P=0^+$ and $0^-$. The number of modes have a branching ratio between 0\% and 
1\%, between 1\% and 2\%, etc. is displayed. 
(b)Decay histograms for $D^0$ non-leptonic decays, combining both
$J^P=0^+$ and $0^-$.
}\label{spectrum1}
\end{figure}

\begin{figure}[htb]
\centerline{\psfig{figure=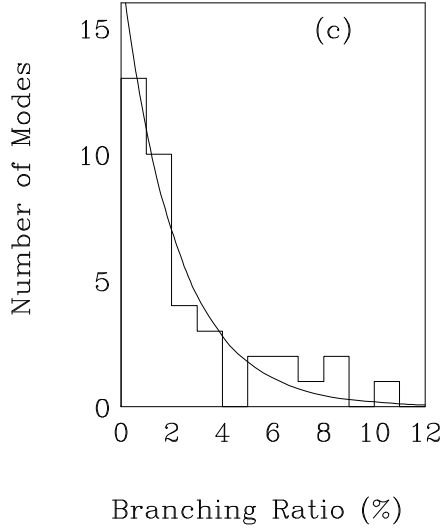,angle=90,width=2.25in}}
\caption[a]{Decay histogram for $D^+$ and $D^0$ non-leptonic decays, combining both
$J^P=0^+$ and $0^-$.  The curve has an exponential fall-off with the standard
deviation equal to the mean, as predicted by the purely random phase shift 
model. 
}\label{spectrum2}
\end{figure}

The inclusive non-leptonic branching ratios for neutral and charged D mesons are
about 0.8 and 0.6 respectively.  If we ascribe half of each total to $0^-$ 
and half to $0^+$, then for the charged D's, we find $N_{3/2}\approx 8$ for the 
parity-even states and $N_{3/2}\approx 13$ for the parity-odd states.  In both cases,
the final states have $I=3/2$.  The neutral D decays have the added complexity
that they involve both $I=1/2$ and $I=3/2$.  Thus we infer $N_{3/2}+N_{1/2}
\approx 0.4/0.02 \approx 20$.  Roughly, then, $N_{1/2}$ and $N_{3/2}$ are each
about 10, for both the parity-even and parity-odd sectors.  In Fig. 
\ref{spectrum1} we show the distribution of branching ratios for $D^+$ and
$D^0$.  In Fig. \ref{spectrum2}, the data are combined and compared with an
exponential distribution of the form $\exp(-|\amp_b^2|/|{<|\amp_b^2|>})$.  
This form has the virtue of satisfying Eq.(\ref{eq518}), relating the standard
deviation of the distribution of $|\amp_b|^2$ to its mean.
The agreement with this empirical form is generally satisfactory.  If we compute
the standard deviation of the branching ratios, combining data from all decay 
channels of neutral and charged $D$s and ignore phase space corrections, we
find

\begeq
\Delta|\amp_b^2|=(1.2\pm 0.2)<|\amp_b^2|>
\endeq
in agreement with our expectation $\Delta|\amp_b^2|=<|\amp_b^2|>$.

\subsection{B decays}

Because of the larger phase space, the number of channels available for B decays
should be substantially larger than for D decays.  This view is supported by
the absence of any dominant exclusive B-decay modes.  We expect the large $N$ 
limit to be very good in B decays, but because of the greater energy available
for decays, weaker final state interactions are anticipated.  We address this 
by restricting the range of $\delta$.  

The value of $\sin^2\delta$ is equal to the ratio of the elastic to the 
total cross section in a fixed partial wave by Eq. (\ref{xsec}), which to order
$(1/N)$ gives

\begeq
\sin^2\delta={\sigma_{el}\over\sigma_{tot}}.
\endeq
Of course experiments are performed with beams, and therefore measure the sum
of many partial waves.  If we make the crude approximation of assuming the
same elasticity for all partial waves, we may estimate $<\sin^2\delta>
\approx 0.2$, on the basis of typical hadronic cross sections evaluated for
a c.m. energy around 5 GeV.  

Consider two different distributions of $\delta$, one flat between 
$-\deltam$ and $\deltam$,

\begeq
{dP\over d\delta}={1\over 2\deltam};\qquad -\deltam<\delta<\deltam
\endeq
and the other one Gaussian

\begeq
{dP\over d\delta}={1\over \sqrt\pi \delta_0}\exp(-\delta^2/\delta_0^2).
\endeq
If we fix $\deltam=1.66$ and $\delta_0=0.715$, then both give 
$<\sin^2\delta>=0.2$.  The flat distribution gives $<\cos\delta>=0.89$, while the 
Gaussian gives  $<\cos\delta>=0.88$.  

We can apply such models to estimate the breakdown of coherence (which is
essential to factorization) caused by
final state interactions.  We can write Eq. (\ref{eq6.3}) as 

\begeqar
&&<|\amp_b + \amp_{b'}|^2>\non
&&\quad = <\cos\delta>^2|X_b+X_{b'}|^2
  +(1-<\cos\delta>^2)[\meanxabsq+\meanxabsqp],
\endeqar
where we have dropped terms of order $1/N$ and taken the distributions of
$\delta$ and $\delta'$ to be identical.  Factorization leads to the prediction
that some branching ratios are much smaller than their isospin relatives, e.g.,
$BR(B^0\to \Dzbar \rho^0) \ll BR(B^0\to \Dmbar\rho^+)$.  Thus for one isospin
variant we have $X_b+X_{b'}\approx 0$, while for another variant we expect
$X_b+X_{b'}$ to be typical of modes generally, i.e. roughly $\sqrt\meanxabsq
+\sqrt\meanxabsqp$.

If we suppose complete cancellation for one mode without final state 
interactions, then the ratio of the decay rate into that mode divided by the 
rate into the isospin variant can be written

\begeq
{\displaystyle{1-<\cos\delta>^2}
\over {1-<\cos\delta>^2+<\cos\delta>^2
  {|X_b+X_{b'}|^2\over \meanxabsq +\meanxabsqp}}},
\endeq
where the amplitudes $X_b$ and $X_{b'}$ are pure isospin amplitudes
for the unsuppressed mode.
If all amplitudes have about the same size, this ratio is simply (using the
approximate
value of $<\cos\delta>$ above)

\begeq
{1-<\cos\delta>^2\over 1+<\cos\delta>^2}\approx 0.13,
\endeq
whereas if the modes in question have branching ratios twice as 
as large as typical modes, the ratio
is
\begeq
{1-<\cos\delta>^2\over 1+3<\cos\delta>^2}\approx 0.07.
\endeq
It is not possible in the case of B decays to estimate the relation between any 
particular amplitude and the mean of the amplitudes as in the case of 
D decays.  The complete mixing of the amplitudes by final state interactions
in the case of D decays makes each one a better representative of the whole 
group. 

The current experimental values for the $B\to {\overline D}\rho$ system
are \cite{rpp}

\begeq
BR(B^0\to \Dzbar \rho^0)<5.5\times 10^{-4}\label{fac1}
\endeq
and
\begeq
BR(B^0\to \Dmbar\rho^+)=(7.8\pm 1.3)\times 10^{-3}\label{fac2}.
\endeq

Final state interactions are not the only contribution to decays like
$B^0\to \Dzbar \rho^0$. The small value of the 
ratio $
BR(B^0\to \Dzbar \rho^0)/ BR(B^0\to \Dmbar\rho^+)$ indicated by
Eqs. (\ref{fac1}) - (\ref{fac2}) suggests that the final state interactions, 
as 
characterized by $<\sin^2\delta>$ are probably not too large.  Of course, even
if there are large final-state interactions, by chance the phases for 
amplitudes contributing to a single final state might end up coherent just as
if there were no final-state interactions.  This may be a case for the
processes $D^0\to K^-\rho^+$ and $D^0\to\kbzero\rho^0$, whose large ratio 
agrees with the factorization prediction.  
However, the failure of factorization in the
$D^0\to K^*\pi$ decays is enough to demonstrate that final-state interactions
are important in $D$ decays, supposing that there is some validity to the factorization
approach.

\clearpage
\appendix
\section{Final-State Interaction Theorem}
\label{appendixa}
\setcounter{equation}{0}
\setcounter{footnote}{0}
\setcounter{table}{0} 

We demonstrate here the final-state interaction theorem in the form appropriate
to our considerations of weak decays.  We begin with an $N$-channel strong 
interaction S-matrix, $S$, and then add a single channel that is coupled to 
the others only by weak interactions.  We write the expanded S-matrix as
\begeq
S^\prime=
\left(
\begin{array}{rr}
1 & U^T\\
U & S
\end{array}
\right),
\endeq
where  $U$ is an $N$-element column vector, and $U^T$ is its 
transpose.  We have assumed T invariance so that the full S-matrix is
symmetric.  Unitarity tells us that

\begeqar
1+U^TU^*=1,\\
U+SU^*=0,\\
SS^\dagger + UU^\dagger =I.
\endeqar

Since $U$ is a weak amplitude and thus small, we drop the $U^2$ terms and are
left with a single relation, $U=-SU^*$.    Because $S$ is not 
just unitary, but symmetric, it can be diagonalized by an orthogonal (rotation)
matrix:

\begeq
S_{diag}=\ortho S \ortho^T,
\endeq
where $S_{diag}$ has diagonal elements $e^{2i\delta_\alpha}; \alpha=1,...N$.  
Now our 
equation is simply

\begeq
\ortho U = -S_{diag}\ortho U^*.
\endeq
If we set $V=\ortho U$, then

\begeq
V_\alpha= - e^{2i\delta_\alpha}V_\alpha^*,
\endeq
and

\begeq
V_\alpha=i e^{i\delta_\alpha} R_\alpha,
\endeq
where $R$ is an arbitrary real number.  The subscript $\alpha$ labels the 
individual eigenchannels of the S matrix.   In particular

\begeq
U_b=i\sum_\alpha\ortho_{\alpha b} e^{i\delta_\alpha} R_\alpha,
\endeq
where we have used Latin subscripts to indicate physical states, as opposed 
to the eigenchannels labeled in Greek.

Thus, in the case of no CP violation (so time reversal invariance is true),
final state interactions provide phases as follows.  Each eigenchannel of the 
S-matrix gets some phase.  The physical states are combinations of the 
eigenchannels and pick up corresponding phases.

How is this changed by CP violation?  We no longer can assume the full S-matrix
is symmetric, but we can assume the strong interaction S-matrix is symmetric.
We write in this instance

\begeq
S^\prime=\left(
\begin{array}{rr}
1 & U_-^T\\
U_+ & S
\end{array}
\right).
\endeq

The idea is that the weak amplitude has weak phases, which come in with one
sign in $U_-$ and the other in $U_+$.  With this convention, $U_+$ gives the 
matrix elements for the decay of the single-particle state, while $U_-$ gives
the matrix elements for the multiparticle states to reform the single-particle
state, or for the time-reversal of the single-particle decay.  Since CP is 
equivalent to T, it is clear that $U_-$ gives the amplitudes for the 
CP-conjugate decays.
Now we find from unitarity

\begeqar
U_++SU_-^*=0,\\
U_-+SU_+^*=0,\\
\endeqar
or equivalently

\begeqar
U_+ + U_-&=& -S(U_+ + U_-)^*,\\
U_+ - U_-&=&  S(U_+ - U_-)^*.\\
\endeqar

Proceeding analogously to our earlier development,

\begeqar
V_\pm&=&\ortho U_\pm,\\
(V_+ + V_-)_\alpha&=& -e^{2i\delta_\alpha}(V_+ + V_-)_\alpha^*,\\
(V_+ - V_-)_\alpha&=& e^{2i\delta_\alpha}(V_+ - V_-)_\alpha^*,\\
\endeqar
and therefore
\begeqar
(V_+ + V_-)_\alpha&=&2ie^{i\delta_\alpha}R_\alpha,\\
(V_+ - V_-)_\alpha&=&-2e^{i\delta_\alpha}I_\alpha,\\
\endeqar
where $R_\alpha,I_\alpha$ are real numbers.  Altogether

\begeqar
V_{\pm \alpha}&=&ie^{i\delta_\alpha}
  (R_\alpha\pm iI_\alpha)\equiv ie^{i\delta_\alpha}
    e^{\pm i\phi_\alpha}W_\alpha,\\
U_{\pm b}&=&i\sum_\alpha\ortho_{\alpha b}e^{i\delta_\alpha}(R_\alpha
              \pm iI_\alpha)\equiv 
              i\sum_\alpha\ortho_{\alpha b}e^{i\delta_\alpha}
                     e^{\pm i\phi_\alpha}W_\alpha.
\endeqar

When an approximate selection rule or dynamical suppression mechanism strongly 
hinders transitions between two groups of states, it is sensible to treat 
transitions between the two groups as a perturbation.  For example, transitions
among the states $\psi\pi\pi$, $D{\overline D}$, and $\pi\pi$ are all 
allowed by strong interactions, but inhibited dynamically.  We can represent
such a state of affairs by an S-matrix that is nearly block diagonal.  Consider
a circumstance in which two different groups having $N_1$ and $N_2$ channels 
are connected only slightly by the strong interactions, so the full S-matrix is

\begeq
S^\prime=
\left(
\begin{array}{rrr}
1 & U_1^T & U_2^T\\
U_1 & \Sone & \epsilon\\
U_2 & \epsilon^T & \Stwo
\end{array}
\right),
\endeq
where we have assumed CP invariance for the present.  We can diagonalize
$\Sone$ and $\Stwo$:

\begeq
\Soned=\ortho^{(1)}\Sone\ortho^{(1)T};
\qquad \Stwod=\ortho^{(2)}\Stwo\ortho^{(2)T}.
\endeq

If we ignore terms of order $\epsilon^2$, the diagonal elements of $\Soned$ and
$\Stwod$ are of the form $e^{i\delta^{(1)}_\alpha}$, $\alpha=1,...N_1$ and 
$e^{i\delta^{(2)}_\beta}$, $\beta=1,...N_2$. In the diagonalizing basis, as a 
result of unitarity the
mixing matrices satisfy

\begeqar
\sum_\gamma \epsilon_{\alpha\gamma}\epsilon^*_{\alpha'\gamma}&=&0;\qquad
 \alpha\neq\alpha',\non
\sum_\gamma \epsilon_{\gamma\beta}\epsilon^*_{\gamma\beta'}&=&0;\qquad
\beta\neq\beta',\non
\endeqar
and they can be written in the form
\begeq
\epsilon_{\alpha\beta}=i{\overline \epsilon}_{\alpha\beta}
e^{i(\delta^{(1)}_\alpha +\delta^{(2)}_\beta)},
\endeq
where ${\overline\epsilon}_{\alpha\beta}$ is real.

Writing $V_i=\ortho^{(i)} U_i$, we find
\begeqar
2 {\rm Re}\ (e^{-i\delta^{(1)}_\alpha}V^{(1)}_\alpha)+i{\overline\epsilon}_{
                \alpha\beta}e^{i\delta^{(2)}_\beta}V^{(2)*}_\beta&=&0,\non
2 {\rm Re}\ (e^{-i\delta^{(2)}_\beta }V^{(2)}_\beta) +i{\overline\epsilon}_{
                \alpha\beta}e^{i\delta^{(1)}_\alpha}V^{(1)*}_\alpha&=&0.\non
\endeqar
The solution is
\begeqar
V^{(1)}_\alpha&=&ie^{i\delta^{(1)}_\alpha} W^{(1)}_\alpha
                        -\half e^{i\delta^{(1)}_\alpha}\sum_\beta{\overline\epsilon}_{
                          \alpha\beta} W^{(2)}_\beta +O(\epsilon^2),\non
V^{(2)}_\beta&=&ie^{i\delta^{(2)}_\beta} W^{(2)}_\beta
                        -\half e^{i\delta^{(2)}_\beta}\sum_\alpha{\overline\epsilon}_{
                          \alpha\beta} W^{(1)}_\alpha +O(\epsilon^2),\non
\endeqar
where $W^{(1)}$ and $W^{(2)}$ are real.  Unitarity provides no information 
about the real part of the amplitude (which here is proceeded by the $i$ that
connects the S matrix with the invariant amplitude).  It is proper, however,
to consider $W^{(i)}$ to have contributions of order $\epsilon$ that arise from
the mixing perturbation.  The reaction rate will thus receive corrections of
order $\epsilon$ and of order $\epsilon^2$ from the real, undetermined part, 
and corrections of order
$\epsilon^2$ from the imaginary part that are explicitly determined.

If CP violation is included, the amplitudes $W^{(i)}$ acquire weak phases that
change sign under CP.  The result is

\begeqar
V^{(1)}_{\pm\alpha}&=&ie^{i\delta^{(1)}_\alpha\pm i\phi^{(1)}_\alpha} 
           W^{(1)}_\alpha
          -\half e^{i\delta^{(1)}_\alpha}\sum_\beta{\overline\epsilon}_{
                          \alpha\beta} e^{\pm i\phi_\beta^{(2)}}W^{(2)}_\beta +O(\epsilon^2)\non
V^{(2)}_{\pm\beta}&=&ie^{i\delta^{(2)}_\beta\pm i\phi^{(2)}_\beta}
                W^{(2)}_\beta
                -\half e^{i\delta^{(2)}_\beta}\sum_\alpha{\overline\epsilon}_{
                          \alpha\beta} 
                e^{\pm i\phi^{(1)}_\alpha}W^{(1)}_\alpha +O(\epsilon^2)\non
\label{mix}
\endeqar

This can be made more concrete by considering a salient example.  The decay
$B\to \pi^+\pi^-$ is important for the measurement of $\sin 2\alpha$, where
$\alpha$ is one of the angles of the unitarity triangle.  The decay receives 
a tree-level contribution from the quark-level process $b\to u\ubar d$, which
has an amplitude proportional to $V_{ub}V_{ud}^*$.  Were that the only 
contribution, there could be no direct CP violation, which requires at least
two weak phases.  On the other hand, CP violation could be observed through
$\Bz\Bzbar$
mixing, and that CP violation would be especially easy to interpret because
it would be unaffected by final state interactions.  However, the $\pi\pi$ 
final state is connected by strong interactions to $D{\overline D}$.  The
decay $B\to D{\overline D}$ proceeds at tree-level through $b\to c\cbar d$, with
an amplitude proportional to $V_{cb}V_{cd}^*$.  Since $V_{ub}/V_{cb}\approx
0.08$, while $V_{cd}/V_{ud}\approx 0.2$, the overall process $B\to 
D{\overline D}\to \pi\pi$ may be important.  This is precisely the process
known colloquially as the ``penguin.''  Of course, the state $D{\overline D}$ 
represents the totality of other similar states, $D{\overline D}\pi,
  D{\overline D}\pi\pi,$ etc.

In the perturbative quark-level calculation, the penguin diagram contributes
a part that is real (aside from the Kobayashi-Maskawa factor), analogous to the
implicit contribution to $W^{(i)}$ above, and an imaginary part\cite{bander} analogous to 
the explicit mixing displayed in Eq. (\ref{mix}).  In the quark-level 
calculation, this absorptive contribution from the penguin diagram is the 
only source of final state interactions.  In our formalism, the final state
interactions are manifested the various phases of the eigenchannels.

\section{Integrals over the Orthogonal Group}

\setcounter{equation}{0}
\setcounter{footnote}{0}
\setcounter{table}{0} 
\label{appendixb}

We demonstrate here the computation of various invariant tensors obtained
by integrating over the $N$ dimensional orthogonal group.  

If we consider the tensor

\begeq
T_{\alpha a \beta b}=\int \mu(dR) \ortho_{\alpha a}\ortho_{\beta b},
\endeq
we see that 
\begeq
\ortho^\prime_{\gamma\alpha}\ortho^\prime_{\delta\beta}T_{\alpha a \beta b}=
T_{\gamma a \delta b}.
\endeq
This follows from the invariance of the measure, $\mu(dR)$: we can shift the
integration variable from $\ortho$ to $\ortho^\prime\ortho$.  A similar 
manipulation shows that
\begeq
T_{\alpha a \beta b}\ortho^\prime_{ac}\ortho^\prime_{bd}=
T_{\alpha c \beta d}.
\endeq
In other words, the tensor $T$ is invariant.  
We assume below that $N>8$.  This eliminates, for integrals of up to eight 
$\ortho$s,
the possibility of invariants formed from the determinant and means that
every tensor must be made from products of Kronecker $\delta$s that combine
either first indices (which are indicated by Greek letters) or second
indices (which are indicated by Roman letters).

For example,

\begeq
\int \mu(dR) \ortho_{\alpha a}\ortho_{\beta b}
\endeq
must be proportional to $\delta_{\alpha\beta}\delta_{ab}$.  Contracting the 
integrand with $\delta_{ab}$ gives unity, so 

\begeq
\int \mu(dR) \ortho_{\alpha a}\ortho_{\beta b}={1\over N}\delta_{\alpha \beta}
\delta_{ab}.
\endeq

Consider next

\begeq
\int \mu(dR) \ortho_{\alpha a}\ortho_{\beta b} \ortho_{\gamma c}
\ortho_{\delta d},
\endeq
which must be constructed from terms like $\delta_{\alpha \beta}
\delta_{\gamma \delta}\delta_{ab}\delta_{cd}$, $\delta_{\alpha \gamma}
\delta_{\beta \delta}\delta_{ab}\delta_{cd}$, etc.  There are precisely 
nine such terms since there are $4!/(2!2!2!)=3$ ways to pick the Greek-index
tensor and an equal number of ways to pick the Roman-index tensor.  With so few
terms it is easy to perform enough contractions to show that

\begeqar
&&\int \mu(dR) \ortho_{\alpha a}\ortho_{\beta b} \ortho_{\gamma c}
\ortho_{\delta d}\non
&&={1\over N(N-1)(N+2)}\left[(N+1)
(\delta_{\alpha\beta }\delta_{\gamma\delta}\delta_{ab}\delta_{cd}
+\delta_{\alpha\gamma}\delta_{\beta \delta}\delta_{ac}\delta_{bd}
+\delta_{\alpha\delta}\delta_{\beta \gamma}\delta_{ad}\delta_{bc})\right.\non
&&
-(
 \delta_{\alpha\beta }\delta_{\gamma\delta}\delta_{ac}\delta_{bd}
+\delta_{\alpha\beta }\delta_{\gamma\delta}\delta_{ad}\delta_{bc}
+\delta_{\alpha\gamma}\delta_{\beta \delta}\delta_{ab}\delta_{cd}\non
&&\left.
+\delta_{\alpha\gamma}\delta_{\beta \delta}\delta_{ad}\delta_{bc}
+\delta_{\alpha\delta}\delta_{\beta \gamma}\delta_{ab}\delta_{cd}
+\delta_{\alpha\delta}\delta_{\beta \gamma}\delta_{ac}\delta_{bd})\right].
\label{four}\endeqar
In particular, for $\alpha\neq\gamma$.

\begeqar
&&\int \mu(dR) \ortho_{\alpha a_1}\ortho_{\alpha a_2} \ortho_{\gamma a_3}
\ortho_{\gamma a_4}\non
&&={1\over N(N-1)(N+2)}\left[(N+1)\delta_{a_1 a_2}\delta_{a_3 a_4}
-\delta_{a_1 a_3}\delta_{a_2 a_4}-\delta_{a_1 a_4}\delta_{a_2 a_3}\right].
\endeqar
A straightforward extension of this technique to calculate the integral of
eight $\ortho$s leads to consideration of $[8!/(2!2!2!2!4!)]^2=105^2=
11,025$ terms, so a different strategy is required.

Consider the class of tensors formed when all the Greek indices are
identical.
Now consider
\begeq
\int \mu(dR)\prod_{i=1}^4 \ortho_{\alpha a_i}= A_{4}(\Dl{a_1}{a_2}\Dl{a_3}{a_4}
           +  \Dl{a_1}{a_3}\Dl{a_2}{a_4}+\Dl{a_1}{a_4}\Dl{a_3}{a_2}).
\endeq
Multiply by $\Dl{a_3}{a_4}$  and find
\begeqar
\int \mu(dR)\prod_{i=1}^2 \ortho_{\alpha a_i}&=&{1\over N}\Dl{a_1}{a_2}\nonumber
\\
&=&A_{4}(N+2)\Dl{a_1}{a_2},
\endeqar
so that $A_{4}=1/[N(N+2)]$.  Using similar definitions and procedures we find
analogously
\begeqar
\int \mu(dR)\prod_{i=1}^6 \ortho_{\alpha a_i}&=&A_6
    \sum_{pairings}\Dl{a_i}{a_j}\Dl{a_k}{a_l} \Dl{a_m}{a_n} 
,\nonumber\\
A_6&=&{1\over N(N+2)(N+4)},\non
\int \mu(dR)\prod_{i=1}^8 \ortho_{\alpha a_i}&=&A_8
    \sum_{pairings}\Dl{a_i}{a_j}\Dl{a_k}{a_l} \Dl{a_m}{a_n} \Dl{a_p}{a_q}
,\nonumber\\
A_8&=&{1\over N(N+2)(N+4)(N+6)},
\endeqar
where the product over pairs has $105=8!/(2!2!2!2!4!)$ terms.

Consider the integral of the product of six $\ortho$s.  The case of six identical
Greek indices is computed above.  There are two other possibilities: 
four identical indices and one distinct identical pair or three distinct pairs.

Let us write $(ij)$ for $\delta_{\alpha_i \alpha_j}$.  Then we can represent
the integral of six $\ortho$s, with four identical Greek indices as
\begeqar
&&\int \mu(dR) \ortho_{\alpha a_1}\ortho_{\alpha a_2}\ortho_{\alpha a_3}
       \ortho_{\alpha a_4}\ortho_{\beta a_5}
                   \ortho_{\beta a_6}\nonumber\\
&&\qquad=A_{42}\left\{(12)(34)(56)+(13)(24)(56)+(14)(23)(56)\right\}\nonumber\\
&&\qquad\qquad +B_{42}\left\{(12)(35)(46)+(12)(36)(45)+(13)(25)(46)+(13)(26)(45)
\right. \nonumber
\\ &&\qquad\qquad\qquad\qquad
           +(14)(25)(36)+(14)(26)(35)  +(15)(23)(46)+(15)(24)(36)\nonumber\\
   &&\left.\qquad\qquad\qquad\qquad\quad
        +(15)(34)(26)+(16)(23)(45)+(16)(24)(35)+(16)(34)(25)\right\},\non
\endeqar
where the twelve $B_{42}$ terms  mix $5$ and $6$ with $1,2,3,4$.  
Contracting the integral over six $\ortho$'s with $(56)$ gives,

\begeqar
&&A_4[(12)(34)+(13)(24)+(14)(23)]\nonumber\\
&&\qquad =A_{42} N[(12)(34)+(13)(24)+(14)(23)]\nonumber\\
&&\qquad\quad +4B_{42}[(12)(34)+(13)(24)+(14)(23)],
\endeqar
so
\begeq
{1\over N(N+2)}=NA_{42}+4B_{42}.
\endeq

Contracting next over $(34)$ gives from Eq.(\ref{four}),
\begeqar
&&{1\over N(N-1)(N+2)}[(N+1)(12)(56)-(15)(26)-(16)(24)]\nonumber\\ &&
\qquad =A_{42}(N+2)(12)(56)
+B_{42}\{2(12)(56)+(4+N)[(15)(26)+(16)(25)]\}.\nonumber\\
\endeqar
Thus 
\begeqar
{{N+1}\over N(N-1)(N+2)}&=&(N+2)A_{42}+2B_{42},\nonumber\\
-{1\over N(N-1)(N+2)}&=&(N+4)B_{42},
\endeqar
so we have
\begeqar
A_{42}&=&{N+3\over N-1}\cdot{1\over N(N+2)(N+4)},\nonumber\\
B_{42}&=&{-1\over (N-1)}\cdot{1\over N(N+2)(N+4)}.
\endeqar

We proceed similarly when there are three distinct pairs of Greek indices:
\begin{eqnarray}
&&\int\mu(dR) \ortho_{\alpha a_1}\ortho_{\alpha a_2}\ortho_{\beta a_3}
       \ortho_{\beta a_4}\ortho_{\gamma a_5}
                   \ortho_{\gamma a_6}\non
&&\quad=
A_{222}\left\{\triple 123456\right\}\nonumber\\
&&\quad +B_{222}\left\{\triple 123546+\triple 123645+\triple 153426\right.\nonumber\\
&&\qquad\qquad \left.+\triple 163425 +\triple 132456 +\triple 142356\right\}\nonumber\\
&&\quad+C_{222}\left\{\triple 132546 +\triple 132645 +\triple 142536\right.\nonumber\\
&&\qquad\qquad+\triple 142635+\triple 152346+\triple 152436\nonumber\\
&&\qquad\qquad+\left. \triple 162345+\triple 162435\right\}.\nonumber
\endeqar

Contracting with $(56)$ and comparing with Eq.(\ref{four})
we find from the coefficients of $\double 1234$ and $\double 1324$
\begeqar
NA_{222}+4B_{222}&=&{N+1\over N(N-1)(N+2)},\nonumber\\
NB_{222}+4C_{222}&=&-{1\over N(N-1)(N+2)}.\nonumber\\
\endeqar

Contracting with $(45)$ gives
\begeqar
A_{222}+(N+1)B_{222}+2C_{222}&=&0,\nonumber\\
2 B_{222}+ (N+2)C_{222}&=&0,\nonumber\\
\endeqar
so that
\begeqar
A_{222}&=&{N^2+3N-2\over N(N-1)(N-2)(N+2)(N+4)},\nonumber\\
B_{222}&=&-{N+2\over N(N-1)(N-2)(N+2)(N+4)},\nonumber\\
C_{222}&=&{2\over N(N-1)(N-2)(N+2)(N+4)}.\nonumber\\
\endeqar

For the integral of eight $\ortho$s, we have five cases, indicated symbolically
by 8, 62, 44, 422, and 2222, where the numbers represent the number of identical 
Greek indices.  The `8' case is done above.  For `62' we write

\begeqar
&&\int\mu(dR) \ortho_{\alpha a_1}\ortho_{\alpha a_2}\ortho_{\alpha a_3}
       \ortho_{\alpha a_4}\ortho_{\alpha a_5}
                   \ortho_{\alpha a_6}\ortho_{\beta a_7} \ortho_{\beta a_8}
\nonumber\\[0.1in]
&&=A_{62}\left\{(78)\left[      	
    \triple 123456 + \triple 123546 + \triple 123645\right.\right.\non
&&\qquad +\triple 132456 + \triple 132546 + \triple 132645\non
&&\qquad +\triple 142356 + \triple 142536 + \triple 142635\non
&&\qquad +\triple 152346 + \triple 152436 + \triple 152634\non
&&\qquad\left.\left.
   +\triple 162345 + \triple 162435 + \triple 162534\right]\right\}\non
&&+B_{62}\left\{
    (17)(28)\left[\double 3456 +\double 3546 +\double 3645\right]\right.\non &&\qquad
\left.  +  (17)(38)\left[\double 2456 +\double 2546 +\double 2645\right]...
             \right\}.\non &&\qquad
\endeqar

First contract with $(78)$.  Then identify the coefficient of $\triple 123456$:

\begeq
NA_{62}+6B_{62}=A_{6}={1\over N(N+2)(N+4)}.
\endeq

Next, contract with $(56)$ and identify the coefficient of $\triple 123478$.

\begeq
(N+4)A_{62}+2B_{62}=A_{42}={N+3\over N-1}\cdot{1\over N(N+2)(N+4)}.
\endeq

It follows that
\begeqar
A_{62}&=&{N+5\over N(N-1)(N+2)(N+4)(N+6)},\nonumber\\
B_{62}&=&-{1\over N(N-1)(N+2)(N+4)(N+6)}.\nonumber\\
\endeqar

Now let us deal with the 2222 computation.

\begin{eqnarray}
&&\int\mu(dR) \ortho_{\alpha a_1}\ortho_{\alpha a_2}\ortho_{\beta a_3}
       \ortho_{\beta a_4}\ortho_{\gamma a_5}
                   \ortho_{\gamma a_6}\ortho_{\delta a_7} \ortho_{\delta a_8}
\non
&&\quad =A_{2222}\left\{\quadd 12345678\right\}\non
&&\quad+B_{2222}\left\{(12)[\triple 354768+\triple 354867+
                            \triple 365748+\triple 365847\right.\non
&&                     \qquad\qquad  +\triple 374658+\triple 374568+
                            \triple 384657+\triple 384567]\non
&&     \qquad              +(34)[\triple 152768+\triple 152867+
                            \triple 165728+\triple 165827      	\non
&&                     \qquad\qquad  +\triple 172658+\triple 172568+
\left.                            \triple 182657+\triple 182567]
\ldots\right\}\non
&&   \quad +C_{2222}\left\{\double 1234[\double 5768+\double 5867]
                 + \double 1256[\double 3748+\double 3847]\right. \non
    &&        \qquad      + \double 1278[\double 5364+\double 5463]
                 + \double 3456[\double 1728+\double 1827]\non
     &&    \qquad\left.     + \double 3478[\double 1526+\double 1625]
                 + \double 5678[\double 1324+\double 1423]  \right\}\non
&&   \quad +D_{2222}\left\{
              [\double 1324+\double 1423][\double 5768+\double 5867]\right.\non
       &&     \qquad  +[\double 1526+\double 1625][\double 3748+\double 3847]\non
       &&   \qquad\left.    +[\double 1728+\double 1827][\double 5364+\double 5463] \right\}
\non
&&   \quad+E_{2222}\left\{
\quadd 13254768+\quadd 13254867+\quadd 13264758+\quadd 13264857\right.
\non &&
+\quadd 13274568+\quadd 13274658+\quadd 13284567+\quadd 13284657
\non &&
+\quadd 14253768+\quadd 14253867+\quadd 14263758+\quadd 14263857
\non &&
+\left.\quadd 14273568+\quadd 14273658+\quadd 14283567+\quadd 14283657
\ldots\right\}.
\non 
\endeqar

\begin{sloppy}
First contract with $(78)$.  From the coefficients of $\triple 123456$,
$\triple 123546$, and $\triple 132546$, respectively, we find
\end{sloppy}

\begeqar
NA_{2222}+6C_{2222}&=&A_{222},\non
4B_{2222}+NC_{2222}+2D_{2222}&=&B_{222}.\non
NB_{2222}+6E_{2222}&=&C_{222}.\non
\endeqar

Contracting, instead, with $(67)$, we find

\begeqar
A_{2222}+4B_{2222}+(N+1)C_{2222}&=&0,\non
(N+2)B_{2222}+2C_{2222}+2E_{2222}&=&0,\non
2B_{2222}+D_{2222}+(N+3)E_{2222}&=&0.\non
\endeqar

The solution is

\begeqar
A_{2222}&=&{(N-2)(N^3+9N^2+19N+3)\over N(N-1)(N+1)(N-2)(N+2)(N-3)(N+4)(N+6)},\non
B_{2222}&=&{2N(N+4)\over N(N-1)(N+1)(N-2)(N+2)(N-3)(N+4)(N+6)},\non
C_{2222}&=&-{N^3+6N^2+3N-6\over N(N-1)(N+1)(N-2)(N+2)(N-3)(N+4)(N+6)},\non
D_{2222}&=&{N^2+5N+18\over N(N-1)(N+1)(N-2)(N+2)(N-3)(N+4)(N+6)},\non
E_{2222}&=&-{5N+6\over N(N-1)(N+1)(N-2)(N+2)(N-3)(N+4)(N+6)}.\non
\endeqar

Turning next to 422, we write

\begin{eqnarray}
&&\int \mu(dR) \ortho_{\alpha a_1}\ortho_{\alpha a_2}\ortho_{\alpha a_3}
       \ortho_{\alpha a_4}\ortho_{\beta a_5}
                   \ortho_{\beta a_6}\ortho_{\gamma a_7} \ortho_{\gamma a_8}
\non
&&\quad =A_{422}\left\{\left[
    \double 1234 +\double1324+\double1423\right]\double 5678\right\}\nonumber
    \\[0.1in]
&&\quad +B_{422}\left\{\left[
    \double 1234 +\double1324+\double1423\right][\double 5768+\double 5867
        ]\right\}\non
&&\quad +C_{422}\left\{
    (78)\left[\triple 152634 +\triple 162534 +\triple 153624 +\triple 163524
\right.\right.
\non
  &&\qquad +\triple 154623 +\triple 164523 +\triple 253614+\triple 263514
\non
&&\qquad\left.\left.
             +\triple 254613 +\triple 264513 +\triple 354612+\triple 364512
         \right]\right\}
\non
&&\qquad+ (56)\left[\triple 172834 +\triple 182734 +\triple 173824 +\triple 183724
\right.
\non
   &&\qquad +\triple 174823 +\triple 184723 +\triple 273814+\triple 283714
\non
&&\qquad\left.\left.
             +\triple 274813 +\triple 284713 +\triple 374812+\triple 384712
         \right]\right\}
\non
&&\qquad +D_{422}\left\{
    (67)\left[\triple 152834 +\triple 182534 +\triple 153824 +\triple 183524
\right.\right.
\non
  &&\qquad +\triple 154823 +\triple 184523 +\triple 253814+\triple 283514
\non
&&\qquad\left.\left.
             +\triple 254813 +\triple 284513 +\triple 354812+\triple 384512
         \right]\right\}
\non
&&\qquad+(57)\left[\triple 162834 +\triple 182634 +\triple 163824 +\triple 183624
\right.
\non
   &&\qquad +\triple 164823 +\triple 184623 +\triple 263814+\triple 283614
\non
&&\qquad
    \left.\left.         +\triple 264813 +\triple 284613 +\triple 364812+\triple 384612
         \right]\ldots\right\}\non
&&\qquad +E_{422}\left\{ 
                   \quadruple 5678 +\quadruple 5687 +\quadruple 5768 \right.
\non
 &&\qquad+\quadruple 5786 +\quadruple 5867 +\quadruple 5876
\non
 &&\qquad+\quadruple 6578 +\quadruple 6587 +\quadruple 6758
\non
 &&\qquad\left.+\quadruple 6785 +\quadruple 6857 +\quadruple 6875
\ldots\right\}.
\end{eqnarray}

If we contract this with $(78)$ we get a six-$\ortho$ integral, which we have 
already calculated.  In particular, if we compare the $(12)(34)(56)$ elements
we find

\begeq
NA_{422}+2B_{422}+4C_{422}=A_{42},
\endeq
while comparing the (12)(35)(46) elements gives

\begeq
NC_{422}+4D_{422}+2E_{422}=B_{42}.
\endeq

On the other hand, contracting with (67) gives $\delta_{\beta\gamma}=0$.  Then
looking at the coefficient of $(12)(34)(58)$ tells us

\begeq
A_{422}+(N+1)B_{422}+4D_{422}=0,
\endeq
while $(12)(35)(48)$ gives
\begeq
2C_{422}+(N+2)D_{422}+2E_{422}=0.
\endeq

Contracting with $(45)$ also gives zero.  The coefficient of $(12)(36)(78)$
yields
\begeq
A_{422}+(N+3)C_{422}+2D_{422}=0,
\endeq
while that of $(12)(37)(68)$ requires
\begeq
B_{422}+C_{422}+(N+4)D_{422}=0.
\endeq

From the last four equations we find
\begeqar
A_{422}&=&-{N^3+8N^2+13N-2\over N^2+7N+14}B_{422},\nonumber\\
C_{422}&=&{N^2+5N+2\over N^2+7N+14}B_{422},\nonumber\\
D_{422}&=&-{2(N+2)\over N^2+7N+14}B_{422},\nonumber\\
E_{422}&=&-{N-2\over N^2+7N+14}B_{422}.\nonumber\\
\endeqar

Combining these with the preceding equation gives
\begeqar
B_{422}&=&-{N^2+7N+14\over N(N-1)(N+1)(N-2)(N+2)(N+4)(N+6)},\nonumber\\
\endeqar
so altogether
\begeqar
A_{422}&=&{N^3+8N^2+13N-2\over  N(N-1)(N+1)(N-2)(N+2)(N+4)(N+6)},\nonumber\\
B_{422}&=&-\left({N^2+7N+14\over  N(N-1)(N+1)(N-2)(N+2)(N+4)(N+6)}\right),\nonumber\\
C_{422}&=&-\left({N^2+5N+2\over  N(N-1)(N+1)(N-2)(N+2)(N+4)(N+6)}\right),\nonumber\\
D_{422}&=&{2(N+2)\over  N(N-1)(N+1)(N-2)(N+2)(N+4)(N+6)},\nonumber\\
E_{422}&=&{N-2\over  N(N-1)(N+1)(N-2)(N+2)(N+4)(N+6)}.\nonumber\\
\endeqar

Finally, consider the case where there are two sets of four identical indices:

\begeqar
&&\int \mu(dR) \ortho_{\alpha a_1}\ortho_{\alpha a_2}\ortho_{\alpha a_3}
       \ortho_{\alpha a_4}\ortho_{\beta a_5}
                   \ortho_{\beta a_6}\ortho_{\beta a_7} \ortho_{\beta a_8}
\label{int44}
\nonumber\\
&& =A_{44}\left\{ (12)(34)(56)(78) + (12)(34)(57)(68)  + {\rm etc.}\right\}\nonumber\\
&& +B_{44}\left\{ (12)(35)(46)(78) + (12)(35)(47)(68)  + {\rm etc.}\right\}\nonumber\\
&& +C_{44}\left\{ (15)(26)(37)(48) + (15)(26)(38)(47)  + {\rm etc.}\right\}
\label{fourfour}
\endeqar

Now contract $(78)$ with the eight-$\ortho$ integral to find $A_{44}$ and $B_{44}$
in terms of $A_{42}$ and $B_{42}$, for 
$\alpha\neq\beta$
\begeqar
&&(N+2)A_{44}[(12)(34)+(13)(24)+(14)(23)](56)\nonumber\\
  &&\qquad     +B_{44}\left\{4[(12)(34)+(13)(24)+(14)(23)](56)\right.\nonumber \\
   &&\qquad\qquad      +(N+4)\left.\left[(12)[(36)(45)+(35)(46)]
               +(13)[(26)(45)+(25)(46)]...\right]\right\}\nonumber\\
   && \qquad\qquad +2 C_{44}[(12)(35)(46)+(12)(36)(45)...]   \nonumber\\
              &&=
A_{42}[(12)(34)(56)+(13)(24)(56)+(14)(23)(56)]\nonumber\\
&&\quad +B_{42}[(12)(35)(46)+(12)(36)(45)+(13)(25)(46)+(13)(26)(45)...],
\endeqar
so
\begeqar
A_{42}&=&(N+2)A_{44}+4B_{44},\nonumber\\
B_{42}&=&(N+4)B_{44}+2C_{44}.
\endeqar

We have two equations for our three unknowns, $A_{44}$, $B_{44}$, $C_{44}$.
Two additional equations can be obtained by contracting 
Eq.~(\ref{int44}) with (45), which gives zero since $\alpha\neq\beta$:
\begeqar
&&[A_{44}+(N+5)B_{44}]
[(12)(36)(78)+(12)(37)(68)+(12)(38)(67)+(13)(26)(78)
\nonumber\\
&&\qquad+(13)(27)(68)+(13)(28)(67)+
   (16)(23)(78)+(17)(23)(68)+(18)(23)(67)] 
\nonumber\\
&&+[3B_{44}+(N+3)C_{44}]
[(16)(27)(38)+(16)(28)(37)+(17)(26)(38)
\nonumber\\
&&\qquad+(17)(28)(36)+(18)(26)(37)+(18)(27)(36)]
=0.
\endeqar
Thus
\begeqar
A_{44}+(N+5)B_{44}=0,\nonumber\\
3B_{44}+(N+3)C_{44}=0.
\endeqar

Combining these with 
\begeq
B_{42}=(N+4)B_{44}+2C_{44},
\endeq
we find
\begeqar
A_{44}&=&{(N+3)(N+5)\over (N-1)N(N+1)(N+2)(N+4)(N+6)},\nonumber\\
B_{44}&=&{-(N+3)\over (N-1)N(N+1)(N+2)(N+4)(N+6)},\nonumber\\
C_{44}&=&{3\over (N-1)N(N+1)(N+2)(N+4)(N+6)}.\nonumber\\
\endeqar

\section{Calculation of Fluctuations}

\setcounter{equation}{0}
\setcounter{footnote}{0}
\setcounter{table}{0} 
\label{appendixc}
We explain here how to calculate fluctuations, which require the evaluation
of 
\begeqar
|\amp_b^2|^2
   &=&    \sum_{a,b,c,d}X_a X_c^* X_d X_e^*\times
\nonumber\\&&\quad\sum_{\alpha,\beta,\gamma,\epsilon}\int \mu(dR)\prod_{\kappa}
         d\delta_\kappa \weight
          \ortho_{\alpha b} e^{i\delta_\alpha}\ortho_{\alpha a}
          \ortho_{\gamma b} e^{-i\delta_\gamma}\ortho_{\gamma c}
          \ortho_{\delta b} e^{i\delta_\delta}\ortho_{\delta d}
          \ortho_{\epsilon b} e^{-i\delta_\epsilon}\ortho_{\epsilon e}.\nonumber\\
\endeqar
 
Using the results of
Appendix~\ref{appendixb}, the contraction of $X_aX_c^*X_dX_e^*$ with
the integrals over the orthogonal group are found to be

Category I:

\begeq 
A_8 [6(\sumxabs)^2+3\sumxsq\sumxstsq+ 48\xbabs\sumxabs +12(X_b^2\sumxstsq
+X_b^{*2}\sumxsq) +24\xbabs^2].
\endeq

Category II:

\begeqar
&&A_{44}[(\sumxabs)^2+4\xbabs\sumxabs +4\xbabs^2]\non
&&+B_{44}[3(\sumxabs)^2+ \sumxsq\sumxstsq+ 36\xbabs\sumxabs + 8(X_b^2\sumxstsq
+X_b^{*2}\sumxsq) + 16\xbabs^2 ]\non
&&+C_{44}[2(\sumxabs)^2+2\sumxsq\sumxstsq+  8\xbabs\sumxabs + 4(X_b^2\sumxstsq
+X_b^{*2}\sumxsq) + 4\xbabs^2] .\non
\endeqar

Category III:

\begeqar
&&A_{44}[               \sumxsq\sumxstsq                 + 2(X_b^2\sumxstsq
+X_b^{*2}\sumxsq) + 4\xbabs^2] \non
&&+B_{44}[ 2(\sumxabs)^2+2\sumxsq\sumxstsq+ 32\xbabs\sumxabs +10(X_b^2\sumxstsq
+X_b^{*2}\sumxsq) +16\xbabs^2 ]\non
&&+C_{44}[4(\sumxabs)^2                  + 16\xbabs\sumxabs                                  
                  + 4\xbabs^2] .\non
\endeqar

Category IV:

\begeqar
&&A_{2222}[\xbabs^2] \non
&&+B_{2222}[ 16\xbabs\sumxabs +4(X_b^2\sumxstsq
+X_b^{*2}\sumxsq) +8\xbabs^2 ]\non
&&+C_{2222}[ 4\xbabs\sumxabs +(X_b^2\sumxstsq
+X_b^{*2}\sumxsq) +6\xbabs^2 ]\non
&&+D_{2222}[2(\sumxabs)^2+ 4\xbabs\sumxabs +(X_b^2\sumxstsq
+X_b^{*2}\sumxsq) +3\xbabs^2+ \sumxsq\sumxstsq ]\non
&&+E_{2222}[4(\sumxabs)^2+ 24\xbabs\sumxabs +6(X_b^2\sumxstsq
+X_b^{*2}\sumxsq) +6\xbabs^2 +2\sumxsq\sumxstsq].\non
\endeqar

Category V:

\begeqar
&&A_{422}[2\xbabs^2 + \xbabs\sumxabs] \non
&&+B_{422}[ 3\xbabs\sumxabs +(\sumxabs)^2+ 2\xbabs^2]\non
&&+C_{422}[ 10\xbabs\sumxabs +3(X_b^2\sumxstsq
+X_b^{*2}\sumxsq) +8\xbabs^2 ]\non
&&+D_{422}[3(\sumxabs)^2+ 26\xbabs\sumxabs +5(X_b^2\sumxstsq
+X_b^{*2}\sumxsq) +8\xbabs^2+ \sumxsq\sumxstsq ]\non
&&+E_{422}[2(\sumxabs)^2+ 8\xbabs\sumxabs +4(X_b^2\sumxstsq
+X_b^{*2}\sumxsq) +4\xbabs^2 +2\sumxsq\sumxstsq].\non
\endeqar

Category VI:

Half the terms have

\begeqar
&&A_{422}[2\xbabs^2 + X_b^2\sumxstsq] \non
&&+B_{422}[ \sumxsq\sumxstsq +2X_b^{*2}\sumxsq+X_b^2\sumxstsq+ 2\xbabs^2]\non
&&+C_{422}[ 12\xbabs\sumxabs+4X_b^2\sumxstsq +8\xbabs^2 ]\non
&&+D_{422}[2(\sumxabs)^2+ 20\xbabs\sumxabs +6X_b^2\sumxstsq
+10X_b^{*2}\sumxsq +8\xbabs^2+ 2\sumxsq\sumxstsq ]\non
&&+E_{422}[4(\sumxabs)^2+ 16\xbabs\sumxabs  +4\xbabs^2 ],\non
\endeqar
while the other half have the complex conjugate result.

Category VII:

Half the terms have
\begeqar
&&A_{62}[(3X_b^{*2}\sumxsq+6\xbabs\sumxabs +6\xbabs^2]\non
&&+B_{62}[6(\sumxabs)^2+ 3\sumxsq\sumxstsq+ 42\xbabs\sumxabs + 12X_b^2\sumxstsq
+9X_b^{*2}\sumxsq + 18\xbabs^2 ],\non
\endeqar
while the other half have the complex conjugate result.

We can use these results to determine the fluctuations and the 
direct CP 
violation, Eq. (\ref{directcp}).  In the latter instance, we see that
the factor $e^{i(\delta_\alpha-\delta_\gamma)}
   (e^{i(\delta_\delta-\delta_\epsilon)}
        -e^{-i(\delta_\delta-\delta_\epsilon)})$ interchanges the second line
of Category II in Table (\ref{t1}) with Category III, and the second and third
lines of Category V with the two lines of 
Category VI.  The result is

\begeqar
&&<\left(|{\amp}_{\overline b}^2|-|\amp_b^2|\right)^2>\non
&&=2\left\{
            (A_{44}+B_{44}-2C_{44})\left[
                     N^4\meanxabsq-N^4\meanxsq\meanxstsq\right.\right.\non
&&\qquad\qquad\left.             +4N^2\xbabs\meanxabsq
- 2N^2(X_b^2\meanxstsq+X_b^{*2}\meanxsq)\right]
                 (1-<\cos 2\delta>^2)\non
&&\quad           +\left[ N^2(A_{422}+3B_{422}-2C_{422}+6D_{422}-8E_{422})
[2\xbabs\meanxabsq - (X_b^2\meanxstsq+X_b^{*2}\meanxsq)]\right.\non
&&\qquad\qquad      \left.\left.
        + N^4(B_{422}+D_{422}-2E_{422}) (\meanxabsq-\meanxsq\meanxstsq)\right]
                 <\cos\delta>^2(1-<\cos 2\delta>)\right\}.
\non
\endeqar

In the limit of large $N$, the most important coefficients are $A_{44}$ and
$A_{422}$, each of which is approximately $N^{-4}$.  Thus for large $N$,
\begeqar
&&<\left(|{\amp}_{\overline b}^2|-|\amp_b^2|\right)^2>\non
&&=2\left\{
            \left(
                     \meanxabsq-\meanxsq\meanxstsq
+{2\over N^2}(2\xbabs\meanxabsq- X_b^2\meanxstsq-X_b^{*2}\meanxsq)\right)
                 (1-<\cos 2\delta>^2)\right.\non
&&\quad     \left.      +{1\over N^2}
(2\xbabs\meanxabsq - X_b^2\meanxstsq-X_b^{*2}\meanxsq)
                 <\cos\delta>^2(1-<\cos 2\delta>)\right\}
.\non
\endeqar

\section{Correlations in the Dyson Model}

\setcounter{equation}{0}
\setcounter{footnote}{0}
\setcounter{table}{0} 
\label{appendixd}
We further define n-level cluster functions, $T_n$, via
\begeqar
&&\sum_{n=0}^\infty\int\cdots\int_0^{2\pi}R_n(\delta_1,...\delta_n)
         A(\delta_1)\cdots 
A(\delta_n)d\delta_1\cdots d\delta_n
\nonumber\\
&&
=\exp
 \sum_{n=0}^\infty{(-1)^{n-1}\over n!}
            \int\cdots\int_0^{2\pi}T_n(\delta_1,...\delta_n)
    A(\delta_1)\cdots A(\delta_n)d\delta_1\cdots d\delta_n,\non
\endeqar
where $A(\delta)$ is an arbitrary function.
In particular,

\begeq
R_2(\delta_1,\delta_2)=T_1(\delta_1) T_1(\delta_2)-T_2(\delta_1,\delta_2).
\endeq

In a tour de force, Dyson calculated $T_2$ and showed how to calculate, in 
principle, the higher-order $T$s.  Explicitly, for even $N=2m$, and with
$p,q=-m+\half,\ldots m-\half$

\begeqar
T_2(\delta,\phi)&=&-\sum_p{p\over 4\pi i}\epsilon(\delta-\phi)
 e^{ip(\delta-\phi)}\nonumber\\
&&+{1\over 8\pi^2}\sum_{pq}\left(2+{p\over q}+{q\over p}\right)e^{i(p-q)(\delta
-\phi)},
\endeqar
where $\epsilon(\zeta)=\zeta/|\zeta|$.

It is clear that 
\begeqar
&&\int_0^{2\pi}\int_0^{2\pi} d\delta d\phi R_2(\delta,\phi)=N(N-1),\nonumber\\
&&\int_0^{2\pi} d\delta T_1(\delta)=N,\nonumber\\
&&\int_0^{2\pi}\int_0^{2\pi} d\delta d\phi T_2(\delta,\phi)=N.\nonumber\\
\endeqar
We need to calculate
\begeq
\int_0^{2\pi}\int_0^{2\pi} d\delta d\phi T_2(\delta,\phi)\cos(\delta-\phi).
\endeq
We first compute, for a half-integer $\nu$
\begeq
\int_0^{2\pi}\int_0^{2\pi} d\delta d\phi 
\epsilon(\delta-\phi)
 e^{i\nu(\delta-\phi)}={4\pi i\over \nu},
\endeq
from which we ascertain

\begeqar
&&\int_0^{2\pi}\int_0^{2\pi} d\theta d\phi 
\sum_p-{p\over 4\pi i}\epsilon(\theta-\phi)
 e^{ip(\theta-\phi)}\cos(\theta-\phi)
\nonumber\\
&&\qquad =-\half\sum_p\left({p\over p+1}+
        {p\over p-1}\right)=-2m+{1\over m+\half}+{1\over m-\half}\nonumber\\
&&\qquad =-N+{2\over N+1}+{2\over N-1}.
\endeqar

For $\mu\neq 0$ an integer, we find
\begeq
\int_0^{2\pi}\int_0^{2\pi} d\theta d\phi 
 e^{i\mu(\theta-\phi)}=0,
\endeq
so the only contributions to the second part of our calculation come from
$p-q\pm 1=0$ and
\begeqar
&&{1\over 8\pi^2}\int_0^{2\pi}\int_0^{2\pi} d\theta d\phi 
\sum_{pq}\left(2+{p\over q}+{q\over p}\right)e^{i(p-q)(\theta
-\phi)}\cos(\theta-\phi)
\nonumber\\
&&={1\over 4}\sum_p [(2+{p\over p+1}+{p+1\over p})+(2+{p\over p-1} +
    {p-1\over p})]\nonumber\\
&&={1\over 4}\sum_p[8+{1\over p-1}-{1\over p+1}]\nonumber\\
&&=2\cdot 2m +{1\over 4}[-{1\over m+\half}-{1\over m -\half}
                            +{1\over -m-\half}+{1\over -m+\half}]
\nonumber\\
&&=2N-[{1\over N+1}+{1\over N-1}].
\endeqar 

Altogether, then,
\begeq
\int_0^{2\pi}\int_0^{2\pi} d\theta d\phi T_2(\theta,\phi)\cos(\theta-\phi)
=N+{1\over N+1}+{1\over N-1}.
\endeq

It follows that 
\begeqar
\int\prod_{k} d\delta_k\Dyson&=&1\nonumber\\
\int\prod_{k} d\delta_k\Dyson \cos(\delta_1-\delta_2)&=&-{1\over N(N-1)}
[N+{1\over N+1}+{1\over N-1}].
\endeqar

Assembling this with our previous calculation gives

\begeq
<|\amp_b|^2>
            = {1\over N+2}
        \left[2|X_b|^2 + N\meanxabsq - {N^2+1\over (N^2-1)(N-1)}(N\xbabs 
                   - N\meanxabsq)\right].
\endeq

Let us now approximate the correlation functions described by Dyson.  The
correlations are of a very short distance, of order $O(1/N)$.  Thus it makes 
sense to use the approximation
\begeq
T_2(\theta_1,\theta_2)={N\over 2\pi}\delta(\theta_1-\theta_2),
\endeq
where the normalization is determined by the relation
\begeq
\int_0^{2\pi} d\theta T_1(\theta)=N.
\endeq

We proceed analogously, writing
\begeqar
&&\int_0^{2\pi}d\theta_1\int_0^{2\pi}d\theta_2\int_0^{2\pi}d\theta_3
R_3(\theta_1,\theta_2,\theta_3)=N(N-1)(N-2),\non
&&\int_0^{2\pi}d\theta_1\int_0^{2\pi}d\theta_2\int_0^{2\pi}d\theta_3
\left[T_3(\theta_1,\theta_2,\theta_3)-T_1(\theta_1)T_2(\theta_2,\theta_3)
-T_1(\theta_2)T_2(\theta_1,\theta_3)\right.\non
&&\qquad -T_1(\theta_3)T_2(\theta_1,\theta_2)\left.
+T_1(\theta_1)T_1(\theta_2)T_1(\theta_3)\right],
\endeqar
from which we conclude
\begeq
T_3(\theta_1,\theta_2,\theta_3)={N\over\pi}\delta(\theta_1-\theta_2)
\delta(\theta_2-\theta_3).
\endeq
Similarly, 
\begeqar
&&R_4 (\theta_1,\theta_2,\theta_3,\theta_4)=N(N-1)(N-2)(N-3)
\non &&\qquad\times
\left[
-T_4(\theta_1,\theta_2,\theta_3,\theta_4)+T_3 T_1 \ldots  +T_2 T_2 \ldots
      -T_2 T_1 T_1 \ldots +T_1 T_1 T_1 T_1\right],\non
\endeqar
where the arguments are implicit.  In this way we determine the approximation

\begeq
T_4 (\theta_1,\theta_2,\theta_3,\theta_4)={3N\over \pi}
   \delta(\theta_1-\theta_2)\delta(\theta_2-\theta_3)\delta(\theta_3-\theta_4).
\endeq

The required integrations over the eigenphases are

\begeqar
&&\int_0^{2\pi}\int_0^{2\pi} d\theta_1 d\theta_2 {R_2(\theta_1,\theta_2)\over
N(N-1)}e^{i(\theta_1-\theta_2)}\non
&&\approx\int_0^{2\pi}\int_0^{2\pi} d\theta_1 d\theta_2 {-T_2(\theta_1,\theta_2)
+T_1(\theta_1)T_1(\theta_2)\over
N(N-1)}e^{i(\theta_1-\theta_2)}\non
&&\approx\int_0^{2\pi}\int_0^{2\pi} d\theta_1 d\theta_2 {-{N\over 2\pi}
\delta(\theta_1-\theta_2)+
({N\over2\pi})^2\over
N(N-1)}e^{i(\theta_1-\theta_2)}\non
&&=-{1\over N-1}.
\endeqar
and
\begeqar
&&\int_0^{2\pi}\int_0^{2\pi}\int_0^{2\pi}
 d\theta_1 d\theta_2 d\theta_3 
{R_3(\theta_1,\theta_2,\theta_3)\over
N(N-1)(N-2)}e^{i(2\theta_1-\theta_2-\theta_3)}\non
&&\approx\int_0^{2\pi}\int_0^{2\pi}\int_0^{2\pi}
 d\theta_1 d\theta_2 d\theta_3 \non
&&\times\left[ {{N\over \pi}
\delta(\theta_1-\theta_2)\delta(\theta_2-\theta_3)
{e^{i(2\theta_1-\theta_2-\theta_3)}\over
N(N-1)(N-2)}}\right]\non
&&={{2N}\over{N(N-1)(N-2)}},\non
\endeqar
and
\begeqar
&&\int_0^{2\pi}\int_0^{2\pi}\int_0^{2\pi}\int_0^{2\pi}
 d\theta_1 d\theta_2 d\theta_3 d\theta_4 
{R_4(\theta_1,\theta_2,\theta_3,\theta_4)\over
N(N-1)(N-2)(N-3)}e^{i(\theta_1-\theta_2+\theta_3-\theta_4)}\non
&&\approx\int_0^{2\pi}\int_0^{2\pi}\int_0^{2\pi}\int_0^{2\pi}
 d\theta_1 d\theta_2 d\theta_3 d\theta_4\non
&&\times\left\{ {-{3N\over \pi}
\delta(\theta_1-\theta_2)\delta(\theta_2-\theta_3)\delta(\theta_3-\theta_4)
}\right.
\non
&&+\left.({N\over 2\pi})^2 [\delta(\theta_1-\theta_2)\delta(\theta_3-\theta_4)
+\delta(\theta_1-\theta_3)\delta(\theta_2-\theta_4)
+\delta(\theta_1-\theta_4)\delta(\theta_2-\theta_3)\}\right]\non
&&\qquad\qquad\times{
e^{i(\theta_1-\theta_2+\theta_3-\theta_4)}\over
N(N-1)(N-2)(N-3)}\non
&&={{-6N+2N^2}\over{N(N-1)(N-2)(N-3)}}.\non
\endeqar
Of course, we have approximated the correlations and should only work to leading 
order in $N$.  We note that in the various tensors, the leading piece is, in 
every case, the $A$ term, which is, to leading order, $1/N^4$.  We collect the
results in Table~\ref{t2}.

\vspace{0.2in}
\begin{table}[htb]\begin{center}
\begin{tabular}{|l|l|l|l|}\hline
&Tensor&Terms&Eigenphase\\ 
&&&Integral\\ \hline
I&$6(\sumxabs)^2+3\sumxsq\sumxstsq+ 48\xbabs\sumxabs $&$N$&$1$\\ 
&$+12(X_b^2\sumxstsq+X_b^{*2}\sumxsq) +24\xbabs^2$&&\\ \hline
II&$(\sumxabs)^2+4\xbabs\sumxabs +4\xbabs^2$&$2N^2$&$1$\\ 
&&&\\ \hline
III&$\sumxsq\sumxstsq+2(X_b^2\sumxstsq+X_b^{*2}\sumxsq)+4\xbabs^2$&$N^2$&
         $-1/N$\\ \hline
IV&$\xbabs^2$&$N^4$& $2/N^2$\\  \hline
V&$2\xbabs^2 + \xbabs\sumxabs$&$4N^3$&$-1/N$\\ \hline
VI&$4\xbabs^2+X_b^2\sumxstsq+X_b^{*2}\sumxsq$&$N^3$&$0$\\ \hline
VII&$3X_b^{*2}\sumxsq+6\xbabs\sumxabs +6\xbabs^2$&$2N(N-1)$&$-1/N$\\ 
 &$$&$2N(N-1)$&\\ \hline
\end{tabular}\end{center}
\caption[a]{Leading terms in $N$ for the fluctuations in the amplitude squared 
in Dyson's model for the various categories.\label{t2}}
\end{table}
Combining these results and taking only the pieces leading $N$ we find

\begeq
<|\amp_b^2|^2>={2\over N^2}(\sumxabs + |X_b|^2)^2
.\endeq  

\vfill\eject

\end{document}